\documentclass[useAMS,usenatbib]{mn2e}
\usepackage[fleqn]{amsmath}
\usepackage{mathtools}
\usepackage{natbib}
\usepackage{graphicx}
\usepackage{mathrsfs}
\usepackage{aas_macros}
\usepackage{amssymb}
\usepackage{multirow}
\usepackage{hyperref}
\usepackage{color}
\newcommand{\Rom}[1]{\uppercase\expandafter{\romannumeral #1}}
\newcommand{\rom}[1]{\lowercase\expandafter{\romannumeral #1}}
\newcommand{\Msun}{\mathrm{M_\odot}}

\title[DRAGONS---UV luminosity functions]{\mbox{Dark-ages reionization \& galaxy formation simulation \Rom{4}:}
UV luminosity functions of high-redshift galaxies}
\author[Liu et al.]{Chuanwu Liu$^1$\thanks{E-mail: chuanwul@student.unimelb.edu.au},
Simon J. Mutch$^1$, P. W. Angel$^1$, Alan R. Duffy$^{2, 1}$, Paul M. Geil$^1$\newauthor
Gregory B. Poole$^1$, Andrei Mesinger$^3$ and J. Stuart B. Wyithe$^1$\\
$^1$School of Physics, University of Melbourne, Parkville, VIC 3010, Australia\\
$^2$Centre for Astrophysics \& Supercomputing, Swinburne University of Technology, PO Box 218, Hawthorn, VIC 3122, Australia\\
$^3$Scuola Normale Superiore, Piazza dei Cavalieri 7, I-56126 Pisa, Italy}

\begin{document}

\date{\today}

\pagerange{\pageref{firstpage}--\pageref{lastpage}} \pubyear{2014}

\maketitle

\label{firstpage}

\begin{abstract}
In this paper we present
calculations of the UV luminosity function from the Dark-ages Reionization And Galaxy-formation Observables from Numerical
Simulations (DRAGONS) project, which combines N-body, semi-analytic and semi-numerical modelling designed to study galaxy
formation during the Epoch of Reionization. Using galaxy formation physics including supernova feedback, the model naturally
reproduces the UV LFs for high-redshift star-forming galaxies from $z{\sim}5$ through to $z{\sim}10$. We investigate the
luminosity--star formation rate (SFR) relation, finding that variable SFR histories of galaxies result in a scatter
around the median relation of $0.1$--$0.3$ dex depending on UV luminosity. We find close agreement between the model and
observationally derived SFR functions. We use our calculated luminosities to investigate the luminosity function below current
detection limits, and the ionizing photon budget for reionization. We predict that the slope of the UV LF remains steep below
current detection limits and becomes flat at $M_\mathrm{UV}{\gtrsim}{-14}$. We find that $48$ ($17$) per cent of the total UV flux at
$z{\sim}6$ ($10$) has been detected above an observational limit of $M_\mathrm{UV}{\sim}{-17}$, and that galaxies
fainter than $M_\mathrm{UV}{\sim}{-17}$ are the main source of ionizing photons for reionization. We investigate the
luminosity--stellar mass relation, and find a correlation for galaxies with $M_\mathrm{UV}{<}{-14}$ that has the form
$M_*{\propto}10^{-0.47M_\mathrm{UV}}$, in good agreement with observations, but which flattens for fainter galaxies.
We determine the luminosity--halo mass relation to be $M_\mathrm{vir}{\propto}10^{-0.35M_\mathrm{UV}}$, 
finding that galaxies with $M_\mathrm{UV}{=}{-20}$ reside in host dark matter haloes of
$10^{11.0\pm 0.1}\Msun$ at $z{\sim}6$, and that this mass decreases towards high redshift.
\end{abstract}


\section{Introduction}
The luminosity function (LF) is one of the most fundamental observables for high-redshift galaxies. The intrinsic UV continuum of
galaxies is determined by young stellar populations, and is therefore thought to be a good tracer for understanding the star-
formation processes in galaxies. As a result, measurement of the UV LF constrains the buildup
of stellar mass in the high-redshift Universe and the role of galaxies in reionization. Successfully describing the UV LF of
high-redshift galaxies is therefore essential for any model aiming to understand galaxy formation at high redshift.

The hierarchical structure formation scenario \citep{1978MNRAS.183..341W} in a dark energy and cold dark matter ($\Lambda$CDM)
universe can simultaneously explain a wide variety of observational features, and is the currently favoured cosmological model.
In this model, galaxies form in the potential wells of dark matter haloes, where accreted, shock heated gas cools and condenses
into star-forming disks \citep{1980MNRAS.193..189F}. Semi-analytic models have had considerable success in reproducing galaxy-
formation observables based on this scenario
\citep{1991ApJ...379...52W, 1993MNRAS.264..201K, 1994MNRAS.271..781C, 1999MNRAS.303..188K, 2000MNRAS.319..168C, 2006MNRAS.365...11C, 2006MNRAS.370..645B, 2008MNRAS.391..481S, 2011MNRAS.412.1828L, 2013MNRAS.428.2001M, 2015arXiv150908473L}.

A typical semi-analytic model consists of numerical and analytic parts:
(\rom{1}) N-body simulations are used to generate the formation and merger history of dark matter haloes;
(\rom{2}) The interplay amongst the baryonic, stellar and gas components are modelled analytically within these dark matter
structures; (\rom{3}) The observable features, such as the luminosities and colours are evaluated by stellar population
synthesis \citep[e.g.][]{2003MNRAS.344.1000B, 2014MNRAS.439..264G}. When compared with hydrodynamical simulations
\citep[e.g.][]{2010MNRAS.402.1536S, 2015MNRAS.446..521S}
semi-analytic models are far less computationally expensive, since they process the dark-matter-dominated large-scale structure
formation and computationally intensive baryon-dominated galaxy formation separately. On the other hand, semi-analytic models
sacrifice a detailed description of the gas physics, and rely on empirical or idealized laws for galaxy formation properties.

Feedback is extremely important for the formation and evolution of galaxies. Supernovae and active galactic nuclei (AGN)
are typically the two dominant modes of feedback incorporated in semi-analytic models \citep[e.g.][]{2006RPPh...69.3101B}. Apart from this
internal feedback within galaxies, the UV background (UVB) radiation also heats the intergalactic medium (IGM) and can
reduce the infall of gas into the shallow potential wells of small haloes. This UV photo-ionization feedback mechanism
delays the completion of reionization and suppresses the baryon content and star-formation rate (SFR) of small
galaxies \citep{2013MNRAS.432L..51S, 2013MNRAS.432.3340S}. Therefore, including the UVB feedback for individual galaxies is
necessary for accurately modelling galaxy formation during the Epoch of Reionization (EoR).

This is the fourth paper in a series describing the Dark-ages Reionization And Galaxy-formation Observables from Numerical
Simulation (DRAGONS\footnote{http://dragons.ph.unimelb.edu.au}) project, which is based on high resolution and high
cadence N-body simulations \emph{Tiamat}
(\citealt[hereafter Paper-\Rom{1}]{2015arXiv151200559P}; \citealt[Paper-\Rom{2}]{2015arXiv151200560A}), and the semi-analytic \textsc{Meraxes}
model of galaxy formation \citep[hereafter Paper-\Rom{3}]{2015arXiv151200562M}, which has been integrated with a semi-numerical algorithm
for ionization structure \citep{2007ApJ...669..663M}. The aim of this paper is to use stellar population synthesis to characterize
the UV LF of high-redshift galaxies from \textsc{Meraxes}, and to study the ionizing photon budget for reionization. The luminosity
dependence of a variety of intrinsic galaxy properties is also studied.

Galaxies at high redshift are too faint to be observed spectroscopically. However, they can be identified using multi-band
photometry and selected using the Lyman-break technique \citep{1996ApJ...462L..17S}. Using this technique, significant progress
has been made in the past two decades in characterizing the observed UV LF of galaxies towards higher redshifts
\citep{1999ApJ...519....1S, 2007ApJ...670..928B, 2010ApJ...709L.133B, 2010ApJ...709L..16O, 2010MNRAS.403..960M, 2011ApJ...737...90B, 2012ApJ...756..164F, 2013ApJ...768..196S, 2013MNRAS.432.2696M, 2014MNRAS.444.2960D, 2014ApJ...786...57S, 2014ApJ...795..126B}.
The most comprehensive UV LF measurements to date at $z{>}4$ were made by \citet{2015ApJ...803...34B} and \citet{2015arXiv150601035B},
based on the assembly of HST datasets including CANDELS, HUDF09, HUDF12, ERS and BORG/HIPPIES programs. The large number (${>}10000$) of
galaxies at $z{\geqslant}4$ provide statistically reliable UV LFs for testing our semi-analytic model of galaxy formation during the
reionization era.

Young galaxies are strong emitters of UV radiation. High-redshift galaxies are therefore thought to be a significant sources
of reionizing photons. However, galaxies above current detection limits (e.g., an absolute UV magnitude
$M_\mathrm{UV}{\sim}{-17}$ at $z{\sim}6$) are not sufficiently numerous to maintain reionization. Rather, empirical studies
and simulations show a galaxy population down to $M_\mathrm{UV}{=}{-13}$ is
required \citep[e.g.][]{2013ApJ...768...71R, 2014MNRAS.443.3435D, 2015ApJ...802L..19R, 2015ApJ...811..140B}.
The faint-end slope of UV LFs is therefore very important since it determines the number of reionizing photons emitted from the
faint galaxies below current detection limits. On the other hand, theoretically, we expect that baryons in very low-mass dark
matter haloes (${\lesssim}10^8\mathrm{M_\odot}$) cannot efficiently cool and form stars, implying that the LFs may have a
truncation at a very faint luminosity.

Observational and numerical studies have investigated the shape of the UV LF at faint luminosities.
\citet{2015ApJ...814...69A} recently obtained LFs down to $M_\mathrm{UV}{=}{-15.25}$ at $z{\sim}7$ behind lensing clusters,
and found that the faint-end slope remains steep. Using the star-formation histories of Local Group dwarf galaxies obtained
from a color-magnitude diagram analysis, \citet{2014ApJ...794L...3W} inferred the LF at $z{\sim}5$ down to $M_\mathrm{UV}{\sim}{-5}$,
and found no truncation. In contrast, using high-resolution cosmological hydrodynamic simulations, \citet{2014MNRAS.442.2560W}
found the slope of LFs at $z{>}7$ is flat at $M_\mathrm{UV}{>}{-12}$. Further, \citet{2015ApJ...807L..12O} found that the slope is flat
for faint luminosities at $z{>}12$ from calculations with a larger simulation volume. The DRAGONS simulation provides a
framework within which we can consider the faint end of the LF within a self-consistent calculation of the reionization history.

This paper is organized as follows. We first summarize our stellar population synthesis modelling as well as the Lyman-break
colour selection criteria in Section \ref{sec:method}. We then show our results for the observed UV LF of selected galaxies in
Section \ref{sec:lf}. In Section \ref{sec:uvflux}, we calculate the fraction of total UV flux above observed luminosity limits.
We study the relation between UV luminosity and properties of galaxies including SFR and galaxy stellar mass, as well as the
mass of dark matter haloes in Section \ref{sec:uv-galaxy}. Finally, in Section \ref{sec:summary}, we present our conclusions.
We employ a standard spatially-flat $\Lambda$CDM cosmology based on \textit{Planck} 2015 data \citep{2015arXiv150201589P}:
$(h, \Omega_{\rm{m}}, \Omega_{\rm{b}}, \Omega_\Lambda, \sigma_8, n_{\rm{s}})=(0.678, 0.308, 0.0484, 0.692, 0.815, 0.968)$.
All magnitudes are presented in the AB system \citep{1983ApJ...266..713O}.

\section{Modelling UV luminosities}\label{sec:method}
The galaxy formation model used in this work is \textsc{Meraxes} (Paper-\Rom{3}), a new semi-analytic model with updated physics
based on \citet{2006MNRAS.365...11C}. \textsc{Meraxes} is implemented on dark matter halo merger trees generated from the
N-body simulation \emph{Tiamat} described in Paper-\Rom{1}. Our fiducial model is based on the \emph{Tiamat}
simulation, which is run in a $67.8h^{-1}\mathrm{Mpc}$ (comoving) cube box including $2160^3$ particles with a particle mass of
$2.64{\times}10^6h^{-1}\Msun$.

\emph{Tiamat} and \textsc{Meraxes} have been designed for studies of reionization. We have performed halo finding
on 100 snapshots at $z{\geqslant}5$, and thus constructed merger trees with a cadence of one snapshot per ${\sim}10^7$ years.
This resolves the dynamical time of galaxy disks at high redshift, and represents a time resolution comparable to the lifetime
of massive stars, allowing us to include time-resolved supernova feedback. The merger trees are ``horizontally" constructed
so that the semi-analytic model computes properties of all galaxies at each consecutive snapshot. This allows us to implement a
self-consistent calculation of feedback from reionization.

Some basic characteristics of the \textsc{Meraxes} semi-analytic model are described below (for a detailed description of
\textsc{Meraxes} see Paper-\Rom{3}):

(\rom{1}) Cooling: Gas infalling into a halo is assumed to be shocked to the virial temperature of the halo
$T{=}3.59{\times}10^5(V_\mathrm{vir}/100\mathrm{km}\,\, \mathrm{s}^{-1})^2\,\,\mathrm{K}$, where $V_\mathrm{vir}$ is the virial
velocity. The hot gas can subsequently cool via a number of mechanisms, with a cooling time at radius $r$ of
\begin{equation}
t_\mathrm{cool}(r)=\frac32\frac{\bar{\mu}m_\mathrm{p}kT}{\rho_\mathrm{hot}(r)\Lambda(T, Z)},
\end{equation}
where $\bar{\mu}m_\mathrm{p}$ is the mean particle mass, $k$ is the Boltzmann constant, $\rho_\mathrm{hot}$ is the hot gas density,
and $\Lambda(T, Z)$ is the cooling function \citep{1993ApJS...88..253S} which depends on both the
temperature and metallicity of the gas. The hot gas density is assumed to have a simple isothermal spherical distribution:
\begin{equation}
\rho_\mathrm{hot} = \frac{m_\mathrm{hot}}{4\pi R_\mathrm{vir}r^2}.
\end{equation}
The cooling radius, $r_\mathrm{cool}$, is defined as the radius at which $t_\mathrm{cool}$ is equal to the dynamical time of the
halo, $t_\mathrm{dyn}{=}R_\mathrm{vir}/V_\mathrm{vir}$ \citep{2006MNRAS.365...11C}. The gas enclosed within $r_\mathrm{cool}$
has sufficient time to cool and flow to the centre. For haloes which have $r_\mathrm{cool}{\geqslant}R_\mathrm{vir}$, hot gas
directly cools into the central regions of haloes. For haloes which have $r_\mathrm{cool}{<}R_\mathrm{vir}$, a quasi-static
hot atmosphere will form. The mass cooling rate is determined by:
\begin{align}
\begin{split}
\dot{m}_\mathrm{cool} &= 4\pi\rho_\mathrm{hot}(r_\mathrm{cool})r_\mathrm{cool}^2\dot{r}_\mathrm{cool}\\
                      &= \frac12m_\mathrm{hot}\frac{r_\mathrm{cool}V_\mathrm{vir}}{R_\mathrm{vir}^2}.
\end{split}
\end{align}

(\rom{2}) Star formation: Cold gas in the central regions of haloes is assumed to settle into a rotationally supported disk with
a size $r_\mathrm{disk}{=}(3\lambda/\sqrt{2})R_\mathrm{vir}$, where $\lambda$ is the spin parameter of the halo
\citep{1998MNRAS.295..319M, 2006MNRAS.365...11C}. Star formation is assumed to occur if the total amount of cold gas in the disk
exceeds a critical value:
\begin{equation}
m_\mathrm{crit} = 2\pi\Sigma_\mathrm{norm}\left(\frac{V_\mathrm{vir}}{\mathrm{km s^{-1}}}\right)
  \left(\frac{r_\mathrm{disk}}{\mathrm{kpc}}\right) 10^6\Msun,
\end{equation}
where $\Sigma_\mathrm{norm}$ is a free parameter in \textsc{Meraxes}. The SFR in the disk is then given by
\begin{equation}
\dot{m}_\ast = \alpha_\mathrm{SF}\frac{m_\mathrm{cold} - m_\mathrm{crit}}{t^\mathrm{disk}_\mathrm{dyn}},
\end{equation}
where $t^\mathrm{disk}_\mathrm{dyn}{=}r_\mathrm{disk}/V_\mathrm{vir}$ is the disk dynamical time and $\alpha_\mathrm{SF}$ is a free parameter
describing the star-formation efficiency. Galaxy mergers can drive strong turbulence in cold gas and trigger
a burst of star formation. The fraction of total cold gas consumed during the burst is \citep{2001MNRAS.320..504S} 
\begin{equation}
e_\mathrm{burst} = \alpha_\mathrm{burst}(m_\mathrm{small}/m_\mathrm{big})^{\gamma_\mathrm{burst}},
\end{equation}
where $m_\mathrm{small}/m_\mathrm{big}$ is the mass ratio of the merging galaxies, and $\alpha_\mathrm{burst} = 0.56$ and
$\gamma_\mathrm{burst} = 0.7$. The assumed initial stellar mass function (IMF) in our
semi-analytic model is a standard \citet{1955ApJ...121..161S} IMF of the form $\phi\propto m^{-2.35}$ in the mass range
$0.1\Msun{\leqslant}m{\leqslant}120\Msun$.

(\rom{3}) Delayed supernova feedback: \textsc{Meraxes} includes internal galaxy feedback from type \Rom{2} supernovae.
Stars with mass greater than $8\Msun$ will end their lives as type \Rom{2} supernovae and release mass and energy.
In many semi-analytic models \citep[e.g. those based on the Millennium Simulation;][]{2005Natur.435..629S},
the separation between each simulation snapshot
is large, and an assumption of instantaneous supernova feedback is used, with energy and mass released as soon as the relevant
stars are formed. However, \emph{Tiamat} has a much higher time resolution of ${\sim}11\mathrm{Myr}$ in order to resolve the shorter
galaxy dynamical time at high redshift. An $8\Msun$ star, which lives ${\sim}40$Myr will therefore explode ${\sim}3$--$4$
snapshots after it formed. For this reason \textsc{Meraxes} implements a delayed supernova feedback scheme, where a supernova
may explode several snapshots after the star-formation episode.

(\rom{4}) UVB photo-evaporation: \textsc{Meraxes} includes UVB photo-suppression feedback, which leads to a reduced baryon
fraction, $f_\mathrm{mod}$, in individual host dark matter haloes relative to the global baryon fraction $f_b{=}\Omega_b/\Omega_m$
\citep{2013MNRAS.432L..51S}:
\begin{equation}
f_\mathrm{mod} (M_\mathrm{vir}) = 2^{-M_\mathrm{crit}/M_\mathrm{vir}},
\end{equation}
where $M_\mathrm{vir}$ is the mass of the halo, $M_\mathrm{crit}$ is the critical halo mass at which $f_\mathrm{mod} = 0.5$:
\begin{equation}
M_\mathrm{crit} = M_0 {J_{21}}^a\left( \frac{1 + z}{10}\right)^b\left[1 - \left(\frac{1 + z}{1 + z_\mathrm{ion}}\right)^c\right]^d,
\end{equation}
where $J_{21}$ is the local ionizing intensity, $z_{ion}$ is the redshift at which the halo was first exposed to the UVB and
$(M_0, a, b, c, d){=}(2.8{\times}10^9\Msun, 0.17, -2.1, 2.0, 2.5)$ are best fit parameters as found by \citet{2013MNRAS.432L..51S}.
\textsc{Meraxes} embeds a modified version of the code \textsc{21cmFAST} \citep{2011MNRAS.411..955M} in order to construct the
ionization field, and to calculate $z_\mathrm{ion}$ and the average UVB intensity, $\langle {J}_{21}\rangle$.

The free parameters in \textsc{Meraxes} were calibrated to replicate the observed stellar mass function at $z{\sim}5$--$7$
\citep{2011ApJ...735L..34G, 2014MNRAS.444.2960D, 2015A&A...575A..96G, 2015arXiv150705636S}, as well as the \textit{Plank} optical
depth to electron scatting measurements \citep{2015arXiv150201589P}.

\subsection{Stellar population synthesis}
Galaxies contain populations of stars with different ages, which form in one or more progenitor galaxies. From a galaxy at a
specific redshift $z_0$, we trace all progenitors in the merger tree and calculate their total SFR at each snapshot redshift
$z_i$ ($z_i{>}z_0$). The stars formed at $z_i$ have an age of
\[
\tau = t - t^\prime = t(z_0) - t(z_i),
\] where $t(z)$ is the age of the Universe at redshift $z$. Through this process, we build a star-formation history
as a function of time, $\Psi(t)$, for the observed galaxy. Because of the short dynamical time at high redshift, the
starburst can result in a rapid change in UV flux during a single snapshot. Rather than begin the burst at the beginning or end of
the snapshot, we therefore interpolate over $10$ timesteps between each snapshot assuming a constant SFR rate. We find that
our results are insensitive to the precise number of sub-steps.

For a `normal' galaxy without a significant active galactic nucleus, the intrinsic (unattenuated) stellar luminosity at the
rest-frame wavelength $\lambda$ is
\begin{equation}
L_\lambda = \int_0^t \Psi (t^\prime) \mathscr{L}_\lambda (t - t^\prime) \mathrm{d}t^\prime,
\end{equation}
where $\mathscr{L}_\lambda(\tau)$ is the luminosity per unit stellar mass of the coeval population with stellar age $\tau$,
and $\Psi$ is the star-formation history.

In this paper, model stellar energy distributions (SEDs) are generated using the public software package
\textsc{STARBURST99} \citep{1999ApJS..123....3L, 2005ApJ...621..695V, 2010ApJS..189..309L, 2014ApJS..212...14L} with a
Salpeter IMF in the mass range $0.1$--$120\Msun$ in order to be consistent with the calculation of SFR in \textsc{Meraxes}.
The Geneva evolutionary tracks with standard mass loss are selected.
The metallicity is set to $Z{=}0.001$ ($0.05\mathrm{Z}_\odot$), which is appropriate during the EoR. Although
\textsc{Meraxes} computes the evolution in metallicity of the interstellar medium, for clarity we have taken the simple approach of a single constant
metallicity value for star formation. We have checked that assuming metallicity values in the range $0.001{<}Z{<}0.008$ does
not significantly affect our results. We do not include nebular components as they would not affect the UV luminosities of
our model galaxies.

\subsection{Lyman-$\alpha$ absorption}
The spectrum of UV radiation from a high-redshift galaxy passing through intergalactic gas clouds which contain neutral hydrogen will show a
series of Ly$\alpha$ absorption lines at wavelengths shorter than $\lambda{=}1216(1+z)$\AA. The fraction of neutral hydrogen
in the IGM grows rapidly towards high redshifts and the Ly$\alpha$ absorption optical depth is observed to significantly increase,
with the SED of high-$z$ star-forming galaxies showing a dropout at $1216(1+z)$\AA. These galaxies are therefore
called Lyman-break galaxies \citep[LBGs,][]{1996ApJ...462L..17S}.

To mimic the LBG selection process, we calculate the Lyman-dropout feature for our semi-analytic galaxies using a model for IGM
absorption. We adopt an effective optical depth of Ly-$\alpha$ absorption at $z < 5.5$ \citep{2006AJ....132..117F},
\begin{equation}\label{eqn:taueff}
\tau_\mathrm{eff} = (0.85 \pm 0.06)\left(\frac{1+z}{5}\right)^{4.3\pm 0.3}.
\end{equation}
\citet{2006AJ....132..117F} found that the evolution of $\tau_\mathrm{eff}$ significantly accelerates at $z_\mathrm{abs}{>}5.5$, with the
effective optical depth
\begin{equation}
\tau_\mathrm{eff} \propto (1 + z)^{10.9}
\end{equation}
at $z{=}5.5$--$6.3$. For simplicity, we adopt this relation for all redshifts at $z{\geqslant}5.5$. Although this extrapolation is
unphysical, the observed Ly-$\alpha$ flux vanishes at $z{>}6$, so that this assumption does not bias the LBG selection.

\subsection{Dust attenuation}
To compare our luminosities with observations, we need to add the effect of dust attenuation. The rest-frame UV
continuum for a galaxy is assumed to have the form
\begin{equation}
f_\lambda \propto \lambda^\beta,
\end{equation}
where $f_\lambda$ is the flux density per wavelength interval and $\beta$ is the power-law index. For high-redshift galaxies,
$\beta$ can be estimated through photometric SED fitting \citep{2012ApJ...754...83B, 2014ApJ...793..115B}.
UV flux can be strongly attenuated by dust grains within galaxies. This is parameterized as 
\begin{equation}\label{eqn:attenuation}
F_o(\lambda) = F_i 10^{-0.4A_\lambda},
\end{equation}
where $F_i$ and $F_o$ are the intrinsic and observed continuum flux densities respectively, and $A_\lambda$ is the change in
magnitude at rest-frame wavelength $\lambda$. The amount of dust attenuation is wavelength dependent with larger optical depths
for shorter wavelengths. Dust attenuation therefore reddens the spectrum by steepening the observed spectral slope.

Dust attenuation can be estimated through a variety of indicators such as emission line ratios (e.g. Balmer series), the slope
of the rest-frame continuum, the ratio between infrared and UV radiation \citep{1999ApJ...521...64M}, the stellar mass of galaxies
\citep{2009ApJ...698L.116P, 2014MNRAS.437.1268H},
and the SFR of galaxies \citep{2006ApJ...644..792R}. In this work, we adopt a luminosity-dependent dust attenuation
model \citep{2012ApJ...754...83B, 2012ApJ...756...14S, 2014ApJ...793..115B} which is summarized below.

Assuming a constant star-formation history, stellar population synthesis shows that galaxies have similar intrinsic UV continuum
slopes \citep[e.g.][]{1995ApJS...96....9L}. The dust-attenuated UV continuum slope, $\beta$, of galaxies will therefore reflect
the amount of dust attenuation. \citet{1999ApJ...521...64M} established a relation between UV dust attenuation and observed
UV continuum $\beta$:
\begin{equation}\label{eqn:meurer1999}
A_{1600} = 4.43 + 1.99\beta,
\end{equation}
where $A_{1600}$ is the dust attenuation at $1600$\AA. This relation is calibrated by comparison with starburst galaxies in the
local Universe, assuming that high-redshift galaxies have the same spectral properties as local galaxies \citep{1999ApJ...521...64M}.

The key observable for determining the dust-attenuation at high redshifts is the value of $\beta$ for high-redshift galaxies.
Observational studies of high-redshift galaxies show that $\beta$ is larger for galaxies with higher redshifts and lower luminosities
\citep{2012ApJ...754...83B, 2014ApJ...793..115B}. \citet{2014ApJ...793..115B} studied this relation using a large sample
(${>}4000$ sources) of galaxies at $z{\sim}4$--$8$. They found a piece-wise linear relation between the mean of $\beta$ and
$M_\mathrm{UV}$ for galaxies at $z{\sim}4$--$6$:
\begin{equation}
\beta =
  \begin{cases}
     \dfrac{\mathrm{d}\beta}{\mathrm{d}M_{\mathrm{UV}}}(M_\mathrm{UV, AB} + 18.8) + \beta_{M_{\mathrm{UV}=-18.8}},\\
       \hfill M_\mathrm{UV, AB} \leqslant -18.8,\\
     -0.08(M_\mathrm{UV, AB} + 18.8) + \beta_{M_{\mathrm{UV}=-18.8}},\\
       \hfill M_\mathrm{UV, AB} > -18.8,\\
  \end{cases}
\end{equation} 
where ${\mathrm{d}\beta}/{\mathrm{d}M_{\mathrm{UV}}}$ and $\beta_{M_{\mathrm{UV}{=}{-18.8}}}$ are from Table 4 of \citet{2014ApJ...793..115B}.
We use this piece-wise relation for our model galaxies at $z{\sim}5$ and $6$.

For galaxies at $z{\sim}7$ and $8$, we use the linear relation
\begin{equation}\label{eqn:linear_beta_luminosity}
\beta = \frac{\mathrm{d}\beta}{\mathrm{d}M_{\mathrm{UV}}}(M_\mathrm{UV, AB} + 19.5) + \beta_{M_{\mathrm{UV}=-19.5}},
\end{equation}
where $\mathrm{d}\beta/\mathrm{d}M_{\mathrm{UV}}$ and $\beta_{M_{\mathrm{UV}{=}{-19.5}}}$ are from Table 3 of \citet{2014ApJ...793..115B}.

Measurements of $\beta$ at $z{\gtrsim}9$ are limited \citep[e.g.][]{2016MNRAS.455..659W}. We assume that the linear mean
$\beta$--luminosity dependence in \mbox{Equation \ref{eqn:linear_beta_luminosity}} remains valid at $z{\sim}9$ and $10$. We estimate
$\beta_{M_{\mathrm{UV}=-19.5}} = -2.19$ and $ -2.16$ for $z{\sim}9$ and $10$ respectively by linearly fitting the observations
\citep{2014ApJ...793..115B} at $z{\sim}4$--$8$. We set ${\mathrm{d}\beta}/{\mathrm{d}M_{\mathrm{UV}}}{=}-0.16$ for $z{\sim}9$ and $10$,
which equals the mean at $z{\sim}4$--$8$. The uncertainty in this relation is large. However, galaxies at $z{>}9$ are usually
faint and dust will not significantly attenuate the UV continuum.

\begin{figure}
    \includegraphics[width=\columnwidth]{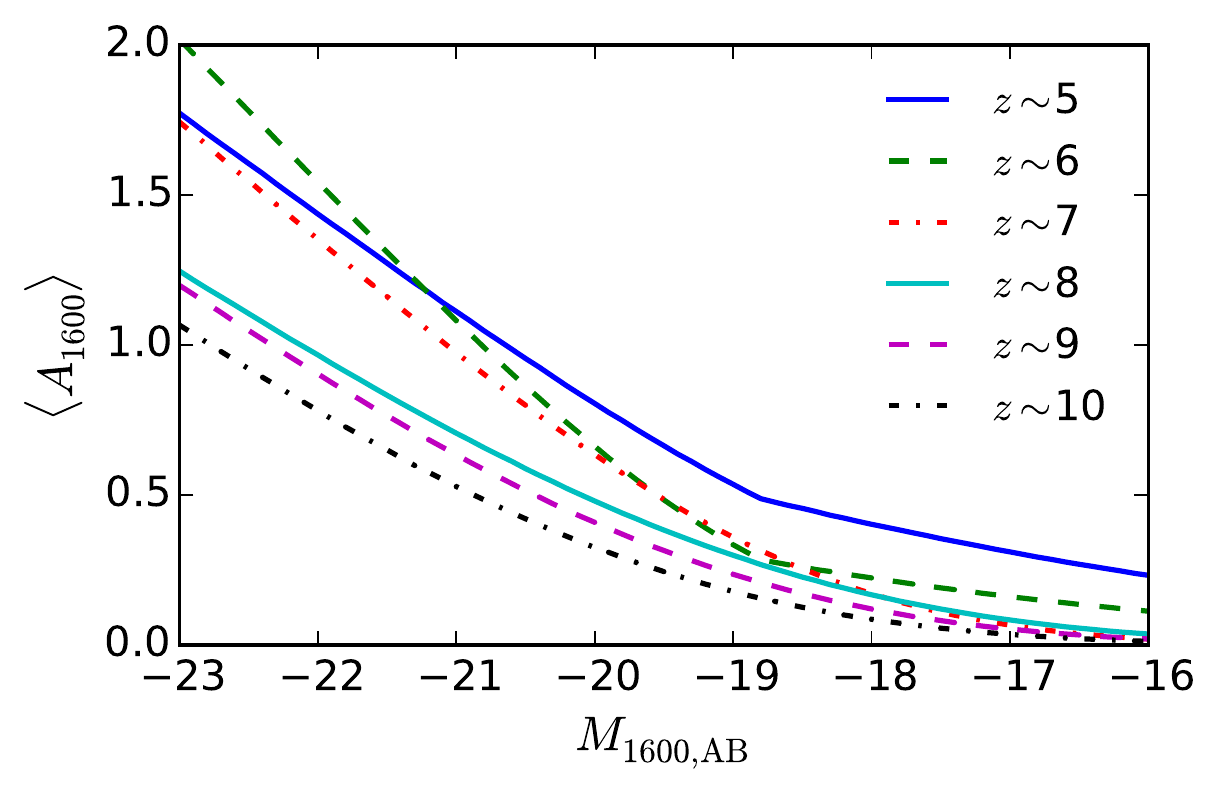}
    \caption{The average dust attenuation at $1600$\AA\, as a function of (dust-attenuated) luminosity at $z{\sim}5$--$10$.
             The linear relation between $A_{1600}$ and the slope of UV continuum $\beta$ from \citet{1999ApJ...521...64M}
             is used. Rest frame observations for $\beta$ are from \citet{2014ApJ...793..115B}.}
\label{fig:a1600}
\end{figure}
\begin{figure*}
    \begin{minipage}{0.99\textwidth}
    \includegraphics[width=\columnwidth]{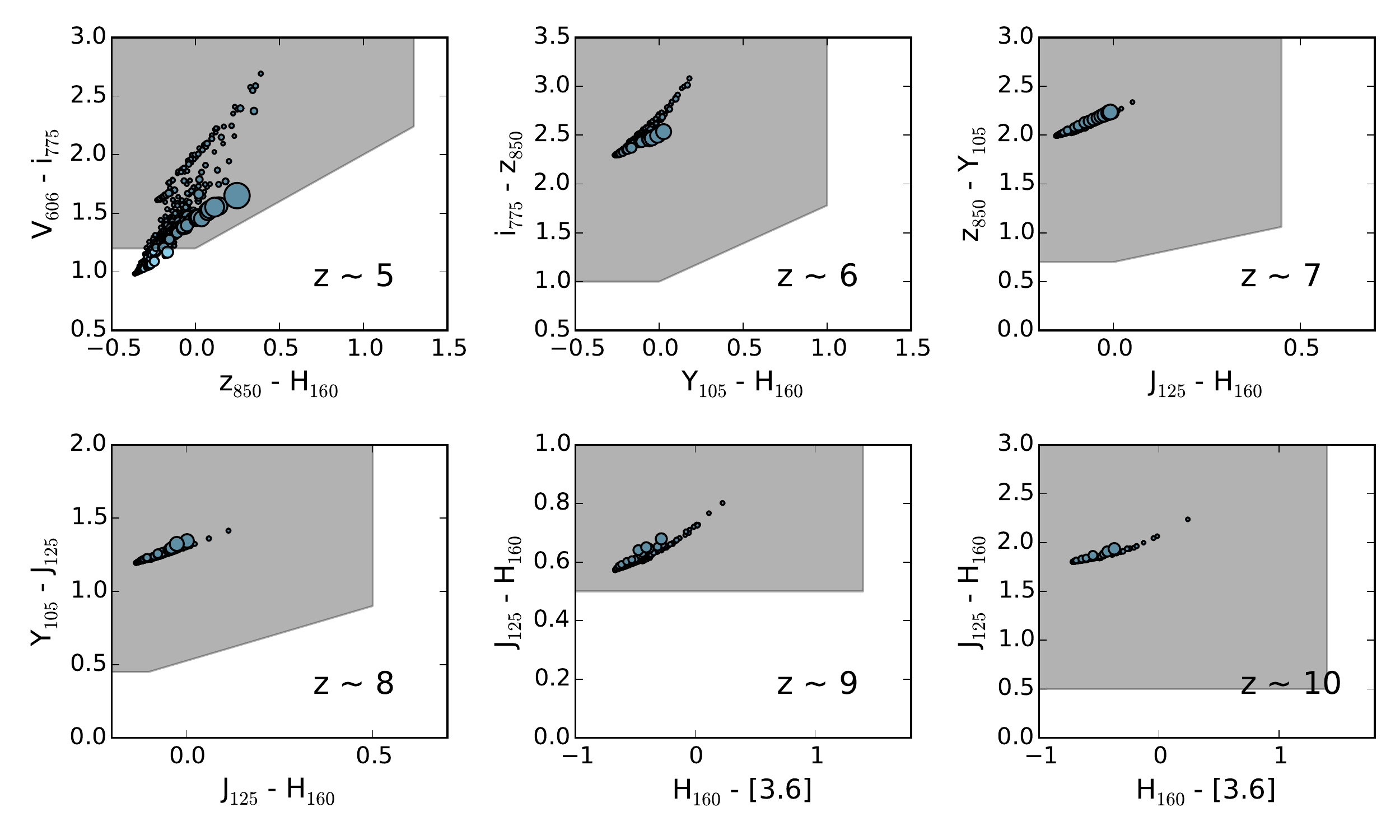}
    \caption{LBG selection criteria used to select star-forming galaxies at $z{\sim}5$--$10$.
         The blue circles show randomly selected galaxies representing 5 per cent of the total sample with $M_{1600}{<}{-15.75}$.
         The areas of circles are proportional to their observed UV luminosities at rest-frame $1600$\AA. The grey-shaded
         regions are the selection regions for LBGs.
         %\hf{We see that all model galaxies can pass the LBG selection except a small fraction of faint galaxies at $z\sim 5$.}
}
\label{fig:colorslt}
\end{minipage}
\end{figure*}
We assume $\beta$ is normally distributed around the mean value with a standard deviation of $\sigma{=}0.35$ \citep{2014ApJ...793..115B}
at all redshifts. From the linear relation in Equation \ref{eqn:meurer1999}, this means that $A_{1600}$ is also normally distributed.
Following \citet{2012ApJ...756...14S}, we set $A_{1600}{=}0$ if $A_{1600}{<}0$ and then calculate the mean, $\langle A_{1600}\rangle$.
Figure \ref{fig:a1600} shows $\langle A_{1600}\rangle$ as a function of observed (dust-attenuated) luminosity for different redshifts.
We can then obtain the relation between $\langle A_{1600}\rangle$ and the intrinsic UV luminosity at $1600$\AA\,using
Equation \ref{eqn:attenuation}. The intrinsic rest-frame magnitude $M_{1500}^i$ and $M_{1600}^i$ are calculated from SEDs using tophat
bands which have a width of $100$\AA\, and centres of $1500$ and $1600$\AA\, respectively.

Dust attenuation in other UV bands can then be estimated using the reddening curve normalized by $\langle A_{1600}\rangle$.
A commonly adopted reddening curve was derived by \citet{2000ApJ...533..682C}\footnote{This can also be approximated by
a simpler reddening curve with extinction optical depth $\tau_\lambda \propto\lambda^{-0.7}$ at
$0.10\mu\mathrm{m}<{\lambda}<0.16\mu\mathrm{m}$ \citep{2000ApJ...539..718C}.}:
\begin{equation}
\label{eqn:redcurve}
k(\lambda) =
  \begin{cases}
    2.659\left(-2.156 + \dfrac{1.509}{\lambda} - \dfrac{0.198}{\lambda^2} + \dfrac{0.011}{\lambda^3}\right) + R_V,\\
       \hfill 0.12\mu \mathrm{m} \leqslant \lambda < 0.63\mu\mathrm{m},\\
    2.659\left(-1.857 + \dfrac{1.040}{\lambda}\right) + R_V,\\
        \hfill 0.63\mu \mathrm{m}\leqslant \lambda \leqslant 2.20\mu\mathrm{m},
   \end{cases}
\end{equation}
where the rest-frame wavelength, $\lambda$, is in units of $\mu$m, $R_V{=}4.05{\pm}0.80$ is the effective obscuration in the $V$
band, and the coefficients are normalized to $E(B{-}V)=k(B){-}k(V)=1$. The change of magnitude due to dust attenuation is
$A_\lambda = E(B{-}V)k(\lambda)$. To obtain the $A_\lambda$ for $\lambda{<}0.12\mu\mathrm{m}$, we extrapolate the reddening curve
$k(\lambda)$ to $\lambda{<}0.12\mu\mathrm{m}$.

\subsection{Lyman-Break selection}\label{sec:selection}
High-redshift galaxies can be selected using multi-band photometric surveys and the Lyman-break technique. To facilitate direct
comparison with observed UV LFs and to study the completeness of LBG selections, we adopt the LBG colour selection criteria
from \citet{2015ApJ...803...34B} to select the model galaxies at $z{\sim}5$--$8$:
\begin{itemize}
  \item The colour selection criterion for $z{\sim}5$ is:
    \begin{multline}
      (V_{606} - i_{775}) > 1.2 \enskip\mathrm{AND}\enskip (z_{855} - H_{160} < 1.3) \\
            \mathrm{AND}\enskip (V_{606} - i_{755} > 0.8(z_{855} - H_{160}) + 1.2).
    \end{multline}
  \item For $z{\sim}6$:
    \begin{multline}
      (i_{775} - z_{850}) > 1.0 \enskip\mathrm{AND}\enskip (Y_{105} - H_{160} < 1.3) \\
           \mathrm{AND}\enskip (i_{775} - z_{850} > 0.78(Y_{105} - H_{160}) + 1.0).
    \end{multline}
  \item For $z{\sim}7$:
    \begin{multline}
      (z_{850} - Y_{105}) > 0.7 \enskip\mathrm{AND}\enskip (J_{125} - H_{160} < 0.45) \\
           \mathrm{AND}\enskip (z_{850} - Y_{105} > 0.8(J_{125} - H_{160}) + 0.7).
    \end{multline}
  \item For $z{\sim}8$:
    \begin{multline}
      (Y_{105} - J_{125}) > 0.45 \enskip\mathrm{AND}\enskip (J_{125} - H_{160} < 0.5) \\
        \mathrm{AND}\enskip (Y_{105} - J_{125} > 0.75(Y_{125} - H_{160}) + 0.525).
    \end{multline}
\end{itemize}
Here $V_{606}$, $i_{775}$, $z_{850}$, $Y_{105}$, $J_{125}$ and $H_{160}$ represent
the magnitudes in ACS and WFC3/IR filter bands F606W, F775W, F850W, F105W, F125W and F160W respectively.
We adopt the colour criteria from \citet{2015arXiv150601035B} to select the galaxies at $z{\sim}9$ and $10$:
\begin{equation}
      (J_{125} - H_{160} > 0.5) \enskip\mathrm{AND}\enskip (H_{160} - [3.6] < 1.4).
\end{equation}
Here $[3.6]$ represents the magnitude in the \emph{Spitzer}/S-CANDELS $3.6\mu\mathrm{m}$ filter.

We calculate the observed dust-attenuated luminosities in these bands for all model galaxies at $z{\sim}5$--$10$, and pass
these through the colour selection criteria. Our model gives the luminosities in dropout bands no matter how faint the galaxies are.
%The additional non-detection criteria in \citet{2015ApJ...803...34B} and \citet{2015arXiv150601035B} are therefore not necessary for our selection.
We do not want to exclude these faint galaxies, so we do not apply the non-detection criteria.
We also do not consider model galaxies at redshifts other than the center of the selection window, and so there is no
contamination by interlopers from other redshifts.

Figure \ref{fig:colorslt} shows the LBG colour-colour selection panels. The galaxies in these panels represent a random sample of
$5$ per cent of the galaxies with $M_{1600}{<}{-15.75}$, which are used for our UV LF determination in Figure \ref{fig:lf}.
The galaxies located in the grey regions are selected as star-forming LBG galaxies.
The size of the circles represents the UV luminosity of galaxies at rest-frame $1600$\AA\,before dust attenuation is applied.

We see that all of our model galaxies are located in the selection regions for $z{\sim}6$--$10$. For $z{\sim}5$, a few faint
galaxies fall outside of the selection region. The UV-bright galaxies have moved toward the upper-right due to dust attenuation
in all panels. Our study shows that if we remove the effect of dust attenuation, a significant number of the brightest galaxies
will be located outside of the selection region at $z{\sim}5$, where LBG selection for the UV-brightest galaxies is very
sensitive to the dust attenuation model.

\section{UV Luminosity functions}\label{sec:lf}
\begin{figure*}
    \begin{minipage}{0.99\textwidth}
    \includegraphics[width=\columnwidth]{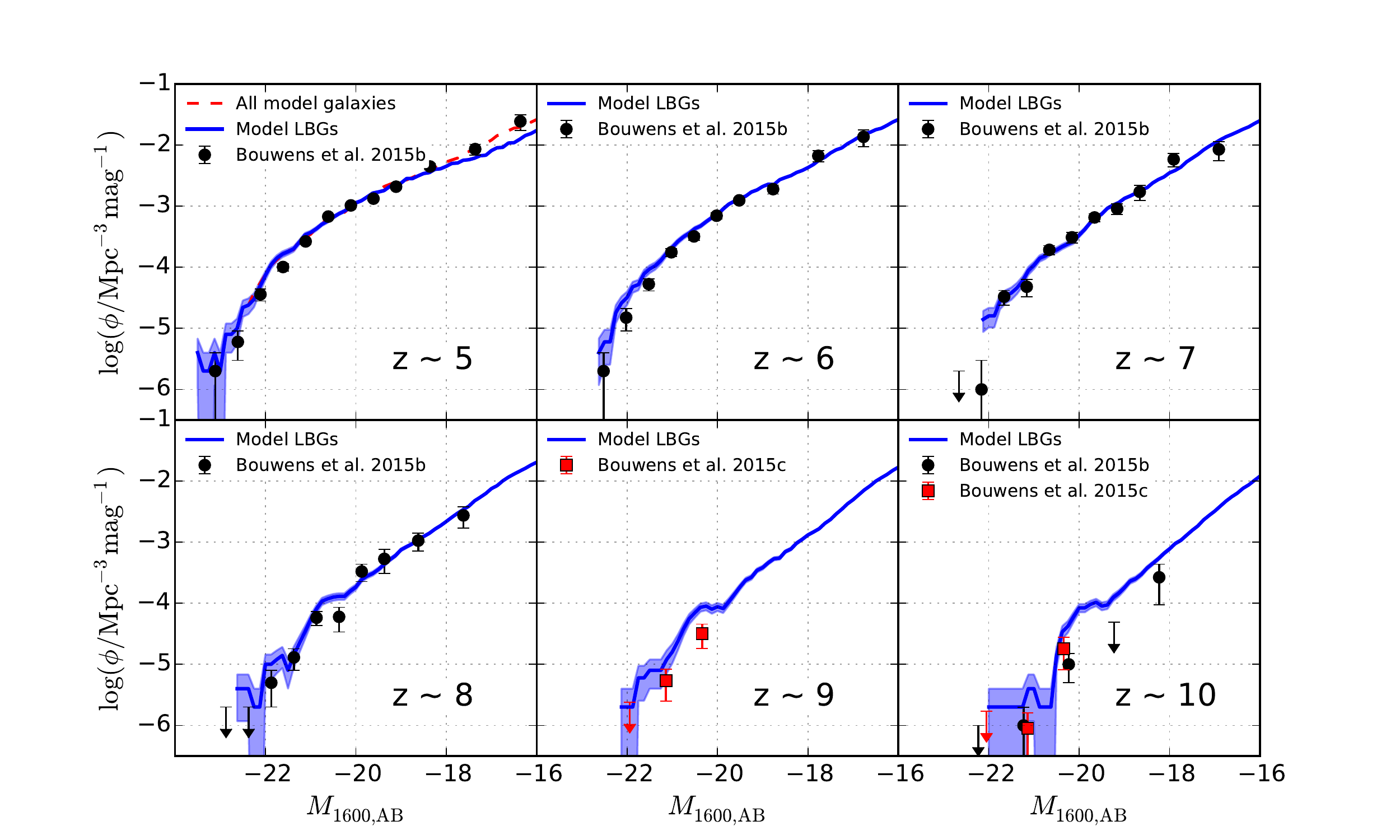}
    \caption{
             Model UV LFs at $z{\sim}5$--$10$ from \textsc{Meraxes}. Blue solid lines show selected LBGs with $1 \sigma$ Poisson
             uncertainties shown as shaded regions. Black circles and red squares are observational data from
             \citet{2015ApJ...803...34B} ($z{\sim}5$, $6$, $7$, $8$ and $10$) and \citet{2015arXiv150601035B}
             ($z{\sim}9$ and $10$) respectively.
             The red dashed line at $z{\sim}5$ shows the UV LF for all model galaxies without the LBG selection. We see close
             agreement between the model and observations.}
    \label{fig:lf}
\end{minipage}
\end{figure*}
Figure \ref{fig:lf} shows our model UV LFs, $\phi(M_\mathrm{UV})$, for LBGs (i.e. the galaxies which passed the LBG selection
criteria) selected at redshifts $z{\sim}5$--$10$. The observed UV LFs are from \citet{2015ApJ...803...34B} at
$z{\sim}5$, $6$, $7$, $8$ and $10$, with additional points from \citet{2015arXiv150601035B} at $z{\sim}9$ and $10$.

For the galaxies at $z{\sim}5$, we also plot the UV LF for all model galaxies for comparison. We see a slight discrepancy between
the LF for all model galaxies and for LBGs, due to the small fraction of faint galaxies with $M_{1600}{>}{-19}$ that do
not pass the LBG selection criteria. For the galaxies at $z{\sim}6$--$10$, all model galaxies with $M_{1600}{<}-15.75$
have passed the LBG selection and are identified as LBGs. Our model, which was calibrated to the stellar mass function at $z{\sim}5$--$7$
(see Paper-\Rom{3}) produces UV LFs at $z{\sim}5$--$10$ that are in excellent agreement with the observations.

\begin{figure*}
    \begin{minipage}{0.99\textwidth}
    \begin{centering}
    \includegraphics[width=0.9\columnwidth]{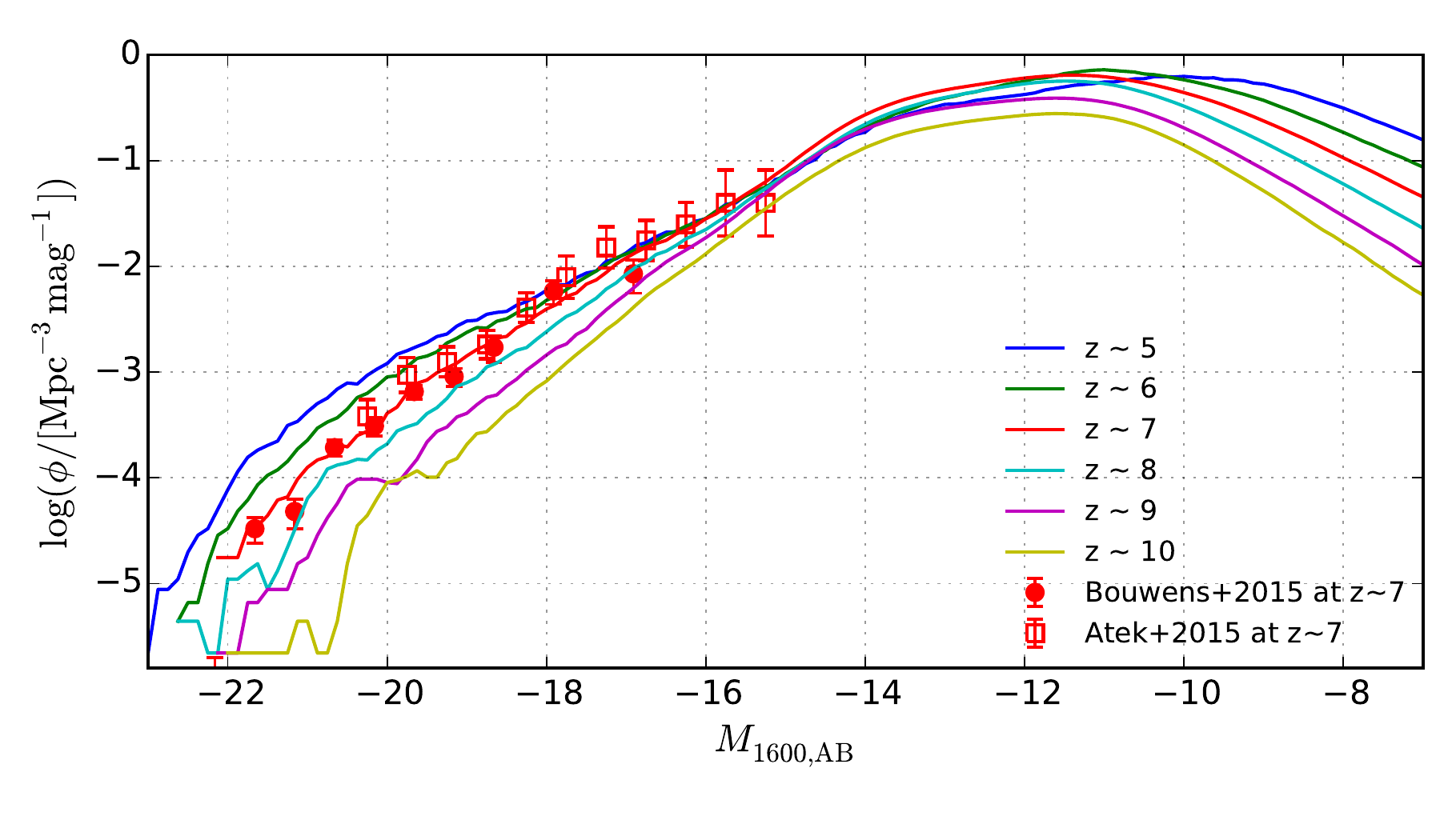}
    \caption{Model UV LFs from \textsc{Meraxes} for all model galaxies extended to low luminosities, illustrating the predicted
             flattening fainter than $M_{1600}{\sim}{-14}$. Red circles and red squares show the observed UV LF from
             \citep{2015ApJ...803...34B} and \citet{2015ApJ...814...69A} at $z{\sim}7$ respectively.}
    \label{fig:lfall}
    \end{centering}
\end{minipage}
\end{figure*}
We can use the model galaxies to study the shape and the evolution of UV LFs to much lower luminosities than observed,
as shown in Figure \ref{fig:lfall}. To exclude the influence of LBG selection criteria, all model galaxies are used hereafter.
The observed LFs from \citep{2015ApJ...803...34B} and
\citet{2015ApJ...814...69A} at $z{\sim}7$ are also shown. \citet{2015ApJ...814...69A} obtained the UV LF down to
$M_\mathrm{UV}{=}{-15.25}$ at $z{\sim}7$. Our prediction is in good agreement with this observation.

We see that the slope of the UV LF remains steep at $M_{1600}{<}{-14}$ and becomes flat at $M_\mathrm{1600}{>}{-14}$.
The UV LFs have a turnover at $M_{1600}{\approx}{-12}$ and then drop towards fainter luminosities. The fact that the faint-end
slope remains steep below current detection limits down to $M_{1600}{=}{-14}$ has important implications for the photon budget
during reionization (see Section \ref{sec:uvflux}). The predicted turnover in the number density of faint galaxies can be traced
to the condition that the halo mass ${\sim}10^8\Msun$ must exceed the hydrogen cooling limit corresponding to a virial
temperature of $10^4$ K before stars can form. A larger value of the cooling mass or temperature will lead to a turnover at
brighter UV magnitude \citep[e.g.][]{2011ApJ...729...99M}. This is also seen by comparing to the relation between the mass
of dark matter haloes and UV luminosity (as discussed in Section \ref{sec:l-mvir}). The flattening at $M_{1600}{>}{-14}$ of LFs
is also a testable prediction of the luminosity below which it becomes likely for the halo masses to drop below the hydrogen
cooling limit.

We also see that the slope of the UV LFs at $M_{1600}{<}{-16}$ steepens towards higher redshift. On the other hand, the
slope at fainter magnitudes ${-16}{<}M_{1600}{<}{-10}$ does not significantly evolve at $z{>}5$. We infer that the implied
continuous growth of extremely faint LFs reflects the ongoing formation of small haloes.
\begin{figure*}
    \begin{minipage}{0.99\textwidth}
    \includegraphics[width=\columnwidth]{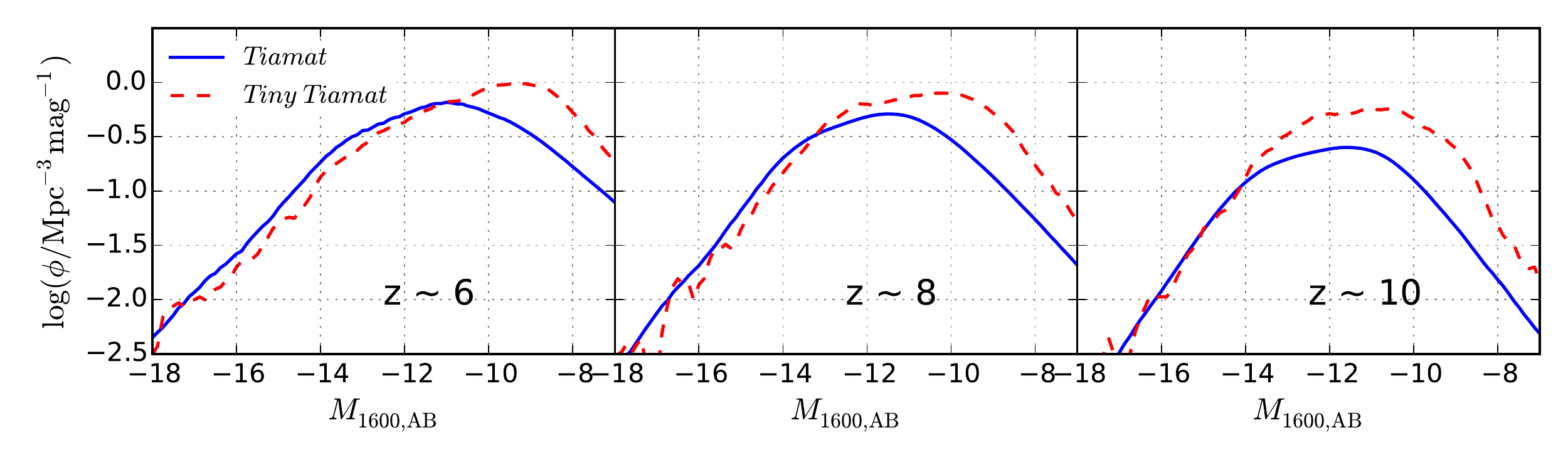}
    \caption{Comparison for UV LFs of all model galaxies based on \emph{Tiamat} and \emph{Tiny Tiamat} simulations at $z{\sim}6$, $8$ and $10$.
             The simulation box sizes are $67.8h^{-1}$Mpc and $10h^{-1}$Mpc, and the particle masses are
             $2.64{\times}10^6h^{-1}\Msun$ and $6.79{\times}10^4h^{-1}\Msun$ for \emph{Tiamat} and \emph{Tiny Tiamat} respectively.
             We see a flattening slope at $M_\mathrm{UV}{\gtrsim}{-14}$ from both simulations.}
\label{fig:lfcomp}
\end{minipage}
\end{figure*}

To investigate how the simulation volume and mass resolution affect the position of the turnover in the UV LF, we compare the
predictions based on the \emph{Tiamat} N-body simulation and the much higher mass resolution \emph{Tiny Tiamat} N-body simulation
(see Paper-\Rom{1}) in Figure \ref{fig:lfcomp}. \emph{Tiny Tiamat} has a $10h^{-1}$ Mpc cubed simulation box and a particle mass
of $6.79{\times}10^4h^{-1}\Msun$, and easily resolves the hydrogen cooling mass at all simulated redshifts. We see that the LFs based
on \emph{Tiamat} and \emph{Tiny Tiamat} generally agree at $M_{1600}{<}{-14}$. However, the model UV LFs based on
\emph{Tiny Tiamat} flatten at $M_\mathrm{1600}{\gtrsim}{-12}$ which is ${\sim}2$ magnitudes fainter than those based on \emph{Tiamat}.
This difference quantifies the combined effects of the hydrogen cooling limit not being completely resolved in the \emph{Tiamat}
simulation until $z{\lesssim}6$, together with merger-triggered star formation in the small haloes near the cooling limit that
cannot be resolved by \emph{Tiamat}.

The flattening slope of high-redshift LFs at $M_\mathrm{UV}{>}{-14}$ has been previously seen in other simulations.
For example, \citet{2015ApJ...807L..12O} carried out a suite of hydrodynamic simulations with an adaptive mesh refinement code
(the \emph{Renaissance Simulations}) which employed a self-consistent radiative transfer reionization scheme and included Population \Rom{3}
stars \citep[see also][]{2014MNRAS.442.2560W, 2012ApJ...745...50W}. These simulations have a dark matter particle of
$2.9\times10^4\Msun$, and show that the $z{>}{12}$ LF flattens at $M_\mathrm{UV}{\gtrsim}{-14}$, in good agreement with our results.

\section{UV flux from galaxies below the detection limit}\label{sec:uvflux}

An important quantity for studies of reionization is the ionizing luminosity emitted by the overall population of galaxies.
There has been extensive discussion in the literature regarding whether enough star formation has been observed to complete
reionization \citep[e.g.][]{2006ARA&A..44..415F, 2010Natur.468...49R, 2013ApJ...768...71R, 2015ApJ...802L..19R, 2015ApJ...811..140B}.
\textsc{Meraxes} describes the stellar mass function of galaxies at $z{\sim}5$--$7$, and the UV LF in the observed range from
$z{\sim}5$--$10$. The predicted UV LFs can be used to calculate the UV luminosity density originating from galaxies above
a luminosity threshold $L_\mathrm{lim}$:
\begin{equation}
\rho_\mathrm{UV} = \int_{L_\mathrm{lim}}^\infty L\phi(L)\mathrm{d}L.
\end{equation}
The emissivity of galaxies (number of ionizing photons emitted into the IGM per second per comoving volume) can then be
estimated using
\begin{equation}
\epsilon = f_\mathrm{esc}\xi_\mathrm{ion}\rho_\mathrm{UV},
\end{equation}
where $\xi_\mathrm{ion}$ is the number of ionizing photons per unit UV luminosity and $f_\mathrm{esc}$ is the fraction of ionizing
photons that escape from the galaxy to ionize the IGM.

\begin{figure*}
    \begin{minipage}{0.99\textwidth}
    \includegraphics[width=\columnwidth]{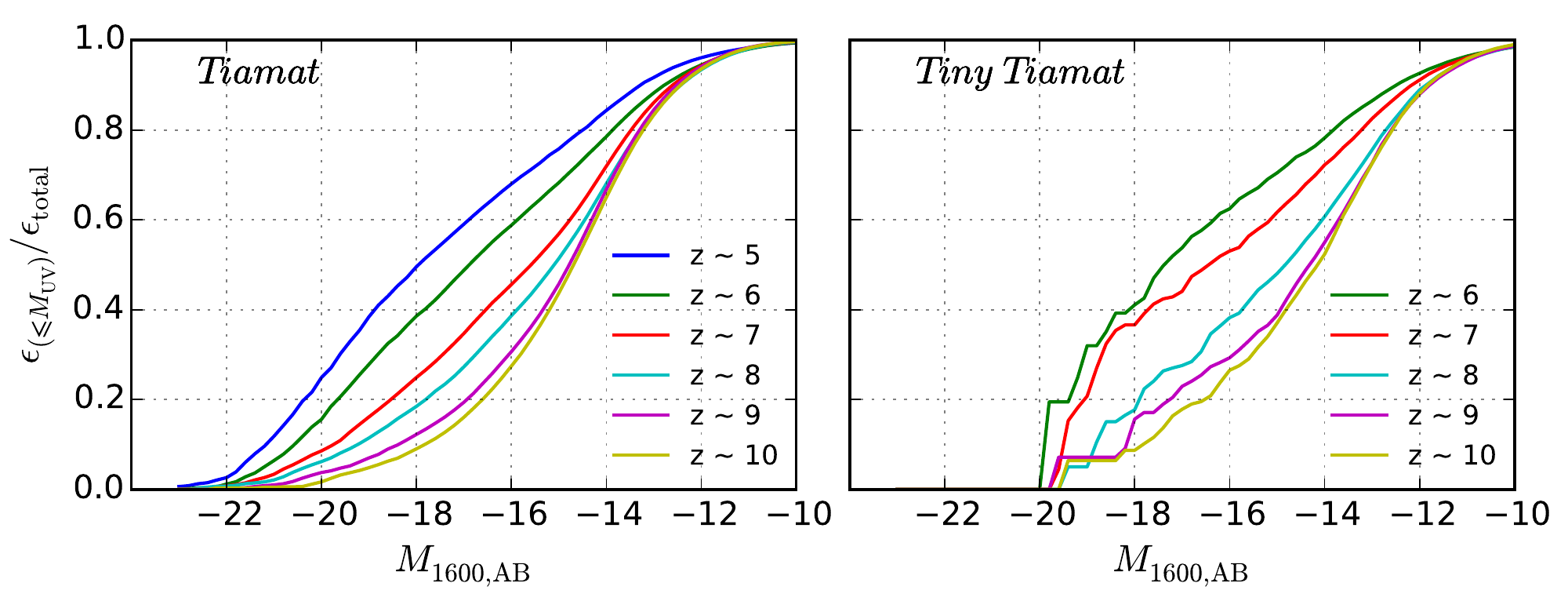}
    \caption{The cumulative fraction of $1600$\AA\, UV flux from the model galaxies brighter the luminosity limit $M_{1600}$
             based on \emph{Tiamat} (left panel) and \emph{Tiny Tiamat} (right panel) $N$-body simulations. We see that
             more than 50 per cent of total UV flux are from galaxies fainter than $M_\mathrm{UV}{=}{-17}$ at $z{>}6$.}
    \label{fig:fluxfrac}
\end{minipage}
\end{figure*}
Figure \ref{fig:fluxfrac} shows the emissivity from galaxies brighter than the limit $M_\mathrm{lim}$ as a fraction of the
total emissivity from all galaxies in the simulation at each redshift:
\begin{equation}
\frac{\epsilon_{(\leqslant M_\mathrm{lim})}}{\epsilon_\mathrm{total}} =
       \frac{\int_{-\infty}^{M_\mathrm{lim}}\phi(M)L(M)\mathrm{d}M}{\int_{-\infty}^{\infty}\phi(M)L(M)\mathrm{d}M}.
\end{equation}
Here we assume $f_\mathrm{esc}$ is constant for all galaxies at each redshift, although it likely scales with halo mass, possibly
compensating for the relative inefficiency in star formation of the faint galaxies
\citep[e.g.][]{2009ApJ...693..984W, 2012MNRAS.423..862K, 2015MNRAS.451.2544P}.
We calculate $\epsilon/\epsilon_\mathrm{total}$ for both simulations based on \emph{Tiamat} and \emph{Tiny Tiamat}. \emph{Tiny Tiamat} misses
the brightest galaxies due to the limited volume, and so presents conservatively low limits on the total fraction of observed
flux. However, we find good agreement between estimates of faint galaxy flux levels among the simulations indicating that our
model based on \emph{Tiamat} is not missing significant luminosity. 

\begin{table}
 \caption{The fraction of UV flux at $1600$\AA\, above the luminosity limits.}\label{tab:fraction}
 \begin{tabular}{llccc}
  \hline
  \parbox{10mm}{}&\parbox{4mm}{$z$}& \parbox{15mm}{$M_{1600}{\leqslant}{-17}$} & \parbox{15mm}{$M_{1600}{\leqslant}{-13}$}&
                                     \parbox{15mm}{$M_{1600}{\leqslant}{-10}$}\\
  \hline
   \multirow{6}{11mm}{\emph{Tiamat}} &
        $5$ & 0.580 & 0.916 & 0.994\\
       &$6$ & 0.478 & 0.884 & 0.995\\
       &$7$ & 0.348 & 0.866 & 0.996\\
       &$8$ & 0.280 & 0.845 & 0.996\\
       &$9$ & 0.202 & 0.850 & 0.997\\
       &$10$ &0.167 & 0.840 & 0.997\\
  \hline
   \multirow{6}{11mm}{\emph{Tiny Tiamat}}
       &$5$ & - & - & -\\
       &$6$ & 0.537 & 0.861 & 0.985\\
       &$7$ & 0.437 & 0.821 & 0.986\\
       &$8$ & 0.277 & 0.756 & 0.986\\
       &$9$ & 0.230 & 0.729 & 0.988\\
       &$10$ &0.177 & 0.725 & 0.990\\
  \hline
 \end{tabular}
\end{table}
The luminosity contributions from galaxies brighter than $M_{1600}{=}{-17}$, $-13$ and $-10$ are shown in Table \ref{tab:fraction}.
Under the assumption of an escape fraction of ionizing radiation that does not depend on mass or redshift, this fraction of
total luminosity equals the fraction of ionizing photons.

In Figure \ref{fig:fluxfrac}, the \emph{Tiny Tiamat}-based simulation shows a truncation of flux at $M_\mathrm{UV}{<}{-20}$,
indicating that simulation volume influences the flux contribution from the brightest galaxies, especially at $z<7$.
However, both simulations give similar values for fractional flux at ${-18}{<}M_\mathrm{UV}{<}{-13}$. We find that the fraction of ionizing
flux from galaxies brighter than the limit $M_\mathrm{UV}{=}{-17}$ evolves continuously from $17$ per cent at $z{\sim}10$ to
$58$ per cent at $z{\sim}5$. This implies that bright galaxies contribute a greater fraction of UV flux at lower redshift
than at high redshift. We see that faint galaxies below a detection limit of $M_\mathrm{UV}{=}{-17}$ at $z{\sim}6$ ($10$)
provide more than 52 (83) per cent of the total flux, and are therefore likely to be the main source of ionizing photons for reionization.
At $z{>}7$, galaxies with luminosities in the range ${-17}{<}M_\mathrm{UV}{<}{-13}$ provide more than ${\sim}50$ per cent of total UV flux.
These results are in agreement with the findings of our hydrodynamic simulations \citep[\emph{Smaug}; see][]{2014MNRAS.443.3435D}.
Due to their inefficient formation, the faintest galaxies ($M_\mathrm{UV}{>}{-10}$) contribute ${<}1$ per cent ionizing flux
at $z{\sim}5$--$10$. Therefore, within the standard model of galaxy formation with a minimum halo mass for star formation
as implemented in \textsc{Meraxes}, the UV flux from these faintest galaxies is negligible during the EoR, and a magnitude
of $M_\mathrm{UV}{\sim}{-10}$ can be considered as an appropriate integration cutoff for luminosity density calculations.

Before leaving this section, we note that although our model successfully reproduces the galaxy UV luminosity functions,
it includes a number of assumptions (e.g., IMF, dust, metallicity and binary populations) which could affect the UV luminosity
of our model galaxies.

\section{UV luminosity-dependence of galaxy properties}\label{sec:uv-galaxy}
In this section, we investigate the relationship between UV luminosity and a series of galaxy properties from
our semi-analytic model. This provides us with predictions for these properties towards very low luminosities.

\subsection{UV luminosity--SFR relation}\label{sec:l-sfr}
\begin{figure*}
    \begin{minipage}{0.99\textwidth}
    \includegraphics[width=\columnwidth]{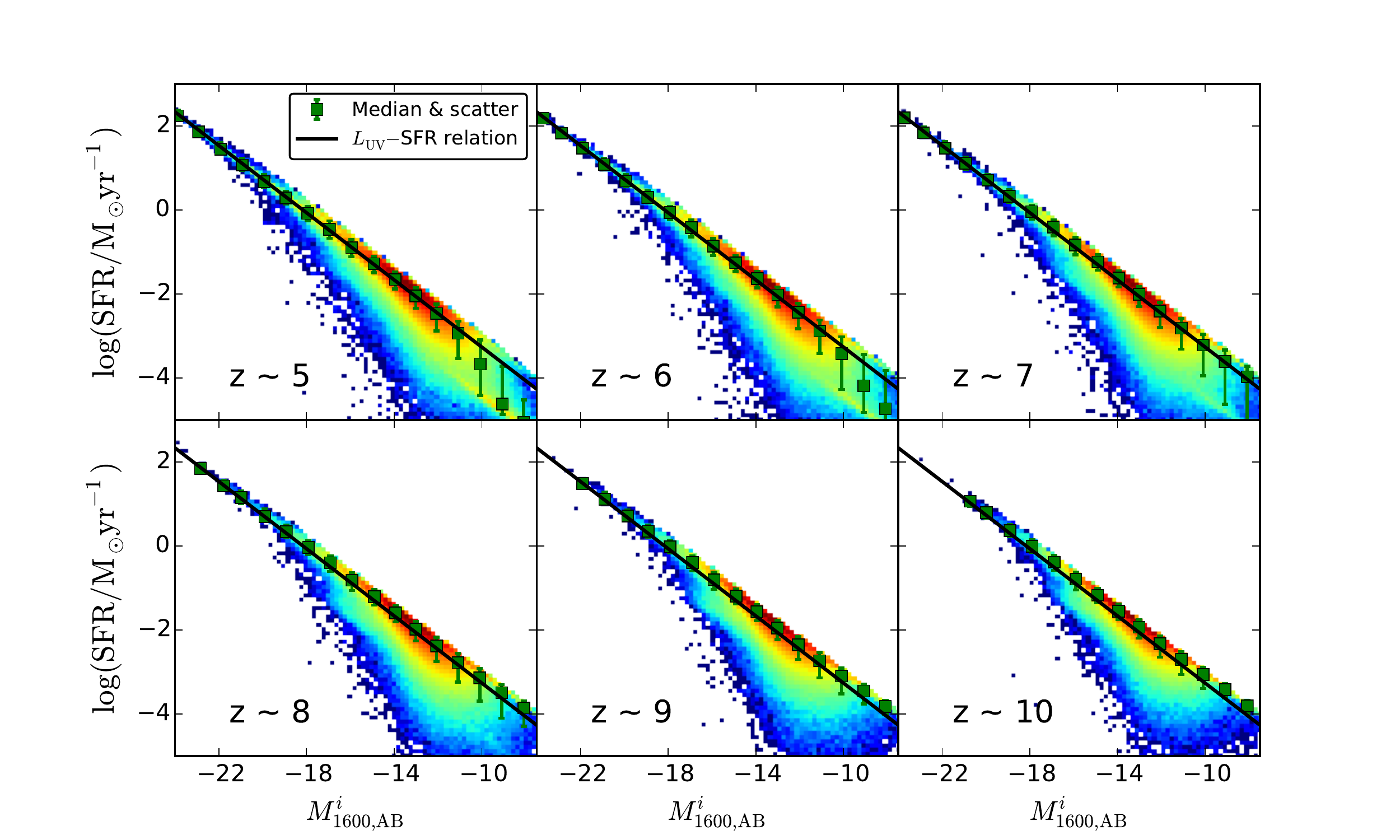}
    \caption{
      The instantaneous SFR of galaxies as a function of their (unattenuated) intrinsic UV luminosity $M_{1600}^i$ at $z{\sim}5$--$10$. The
      colour profile represents the logarithm density of the distribution. The green squares and error bars represent the median
      and 16th to 84th percentiles in bins which contain at least 5 model galaxies
      respectively. The black lines represent the linear relation between $L_\mathrm{UV}$ and SFR shown in
      \mbox{Equation \ref{eqn:l-sfr}} with $\mathcal{K}_\mathrm{UV}{=}1.25\times 10^{-28}$ \citep{1998ARA&A..36..189K, 1998ApJ...498..106M}.
      We see that the Madau-Kennicutt conversion describes the correlation between the UV luminosity and the instantaneous
          SFR from our model well.}
    \label{fig:l-sfr}
    \end{minipage}
\end{figure*}
Since UV flux comes mostly from massive and short-lived stars, the intrinsic UV luminosity is proportional to SFR and
independent of star-formation history over timescales of $t \geqslant t_\mathrm{MS}$, where $t_\mathrm{MS}$ is the
main-sequence time of massive stars \citep{1998ApJ...498..106M}. UV luminosity is therefore thought to be a
good indicator of SFR \citep{1998ARA&A..36..189K}, via the linear relation
\begin{equation}\label{eqn:l-sfr}
\frac{\mathrm{SFR}}{\Msun\mathrm{yr}^{-1}} = \mathcal{K}_\mathrm{UV} \frac{L_\mathrm{UV}}{\mathrm{erg}\cdot \mathrm{s}^{-1} \cdot \mathrm{Hz}^{-1}},
\end{equation}
where the $L_\mathrm{UV}$ is the intrinsic UV luminosity and $\mathcal{K}_\mathrm{UV}$ is a constant. The value of $\mathcal{K}_\mathrm{UV}$ is
model dependent, but can be calibrated via a stellar synthesis model which depends on the IMF, metallicity and star-formation history.
Using a Salpeter IMF in the range $0.1$--$125\Msun$ and an exponential burst of star formation with timescale ${\gtrsim}1$Gyr,
\citet{1998ApJ...498..106M} obtained $\mathcal{K}_\mathrm{UV}{=}1.25{\times}10^{-28}$ for $L_\mathrm{UV}$ in the wavelength range $1500$--$2800$\AA.
Assuming 100Myr of constant star formation and a Salpeter IMF in the range $0.1$--$100\Msun$, \citet{1998ARA&A..36..189K} calibrated
the value to be $\mathcal{K}_\mathrm{UV}{=}1.4{\times}10^{-28}$. \citet{2011MNRAS.411...23W} similarly obtained a value
$\mathcal{K}_\mathrm{UV}{=}1.31{\times}10^{-28}$ using the STARBURST99 population synthesis model employed in this paper.

However, at $z{\gtrsim}6$ the age of the Universe is less than $1$ Gyr, and many galaxies have star-formation histories
shorter than $100$Myr. Therefore, we investigate how well UV luminosity traces the instantaneous SFR using the variable
star-formation histories from our model.

Figure \ref{fig:l-sfr} shows the relation between the intrinsic UV luminosity and the
instantaneous SFR for all galaxies in \textsc{Meraxes} at $z{\sim}5$--$10$. The distribution of SFRs at fixed stellar mass for
the galaxies in these plots indicates the effect of variations in star-formation histories. The distribution shows a sharp upper
limit, corresponding to the youngest galaxies which formed in the latest snapshot. We see that the model predicts a linear relationship
between the UV luminosity and instantaneous SFR, which can be fitted well by the \citet{1998ApJ...498..106M} and
\citet{1998ARA&A..36..189K} relation. The model galaxy SFRs are distributed with a scatter around the median of $\log(\mathrm{SFR})$
varying from \mbox{$\sigma{\sim}0.2$ dex} at $M_{1600}^i{<}{-18}$ to
\mbox{${\sim}0.3$ dex} at $M_{1600}^i{=}{-14}$. The scatter is larger for UV-faint galaxies than for UV-bright galaxies. The
distribution around the median $L$-SFR relation is not log-normal at faint luminosities in our model owing to the minimum star-formation
timescale set by the finite temporal spacing of our simulation snapshots which cuts off the distribution at high SFR.
We fit the $L$-SFR relation using Equation \ref{eqn:l-sfr} for model galaxies brighter than $M_\mathrm{UV}{=}{-14}$ and obtain
$\mathcal{K}_\mathrm{UV}{=}(1.13, 1.16, 1.17, 1.19, 1.24, 1.39)\times 10^{-28}$ at $z{\sim}(5, 6, 7, 8, 9, 10)$. The
value of $\mathcal{K}_\mathrm{UV}$ slightly increases towards higher redsfhit due to the shorter galaxy-formation history and the higher
fraction of young stars at higher redshift.

\subsection{The SFR functions}
\begin{figure*}
    \begin{minipage}{0.99\textwidth}
    \includegraphics[width=\columnwidth]{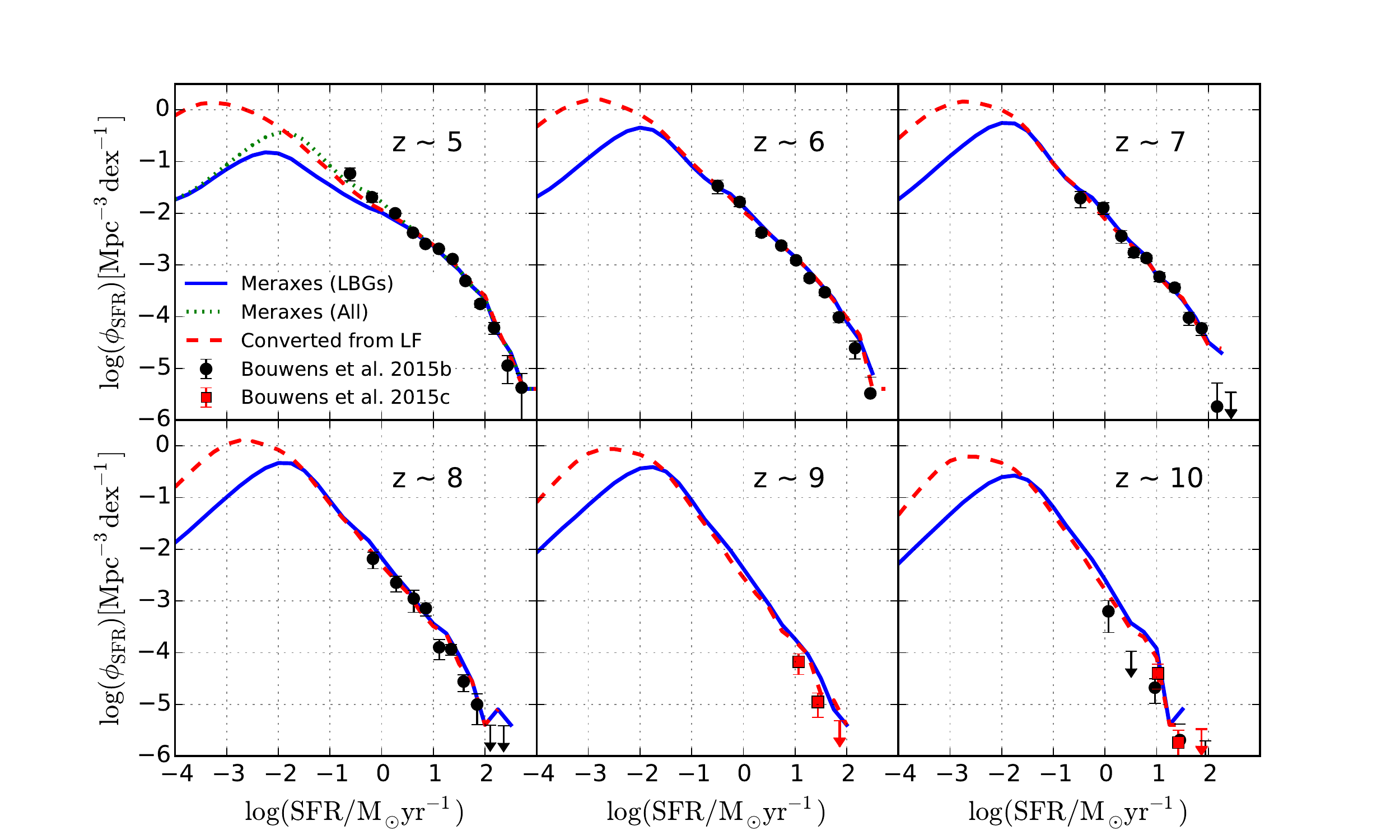}
    \caption{\label{fig:sfrf}
    SFR functions of galaxies at $z{\sim}5$--$10$. The blue lines show the SFR function for LBGs obtained directly from the
    semi-analytic model.
    The dotted green line is the SFR function obtained from the semi-analytic model but without LBG selection at $z{\sim}5$.
    The red dashed lines show the model SFR converted from UV luminosity using the luminosity-SFR relation with
    $\mathcal{K}_\mathrm{UV} = 1.25{\times}10^{-28}$ \citep{1998ARA&A..36..189K, 1998ApJ...498..106M}. Using this relation,
    black circles show the SFR functions converted from the observed UV LFs from \citet{2015ApJ...803...34B} with dust correction
    ($z{\sim}5$, $6$, $7$, $8$ and $10$). Red squares show the SFR functions converted from the observed UV LFs from \citet{2015arXiv150601035B}
    ($z{\sim}9$ and $10$) with dust correction applied. We see close agreement between the SFR functions calculated directly from
    the model and converted from the predicted UV luminosities.}
\end{minipage}
\end{figure*}
An important quantity related to the buildup of stellar mass during reionization is the SFR function \citep[e.g.][]{2012ApJ...756...14S}
which is shown in Figure \ref{fig:sfrf}. The observed SFR function is estimated by converting the UV LF
from \citet{2015ApJ...803...34B, 2015arXiv150601035B} using a \citet{1998ARA&A..36..189K} and \citet{1998ApJ...498..106M} relation.
It is therefore important to investigate whether this assumed conversion introduces bias, and whether the scatter in the
relationship effects the determination of the SFR function. We use two methods to derive the model SFR function:
(\rom{1}) we calculate the predicted instantaneous SFR function using the SFR calculated directly from the semi-analytic model;
and (\rom{2}) to mimic observations, we convert the model UV luminosities to SFRs using Equation \ref{eqn:l-sfr} with
$\mathcal{K}_\mathrm{UV}{=}1.25{\times} 10^{-28}$. The LBG selection is implemented for both of the above methods. Figure \ref{fig:sfrf} shows the
derived model SFR functions together with the observational estimates.

There is close agreement between the predicted model SFR functions and UV-derived model SFR functions
at $\log(\mathrm{SFR}/\Msun\mathrm{yr}^{-1}){>}{-2}$, which in turn agree well with the observational estimates. The small
difference between the predicted SFR function for LBGs and the UV derived SFR function at $z{\sim}5$ is caused by the LBG colour
selection criteria. However, the differences between model-predicted and UV-derived SFR functions at very low SFRs of
$\log(\mathrm{SFR}/\Msun\mathrm{yr}^{-1}){<}{-2}$ show that the estimate of the SFR function using the \citet{1998ApJ...498..106M} and
\citet{1998ARA&A..36..189K} conversion between UV luminosity and SFR will be biased by the scatter of the luminosity--SFR
distribution.

\subsection{UV luminosity--stellar mass relation}\label{sec:l-mg}
\begin{figure*}
    \begin{minipage}{0.99\textwidth}
    \includegraphics[width=\columnwidth]{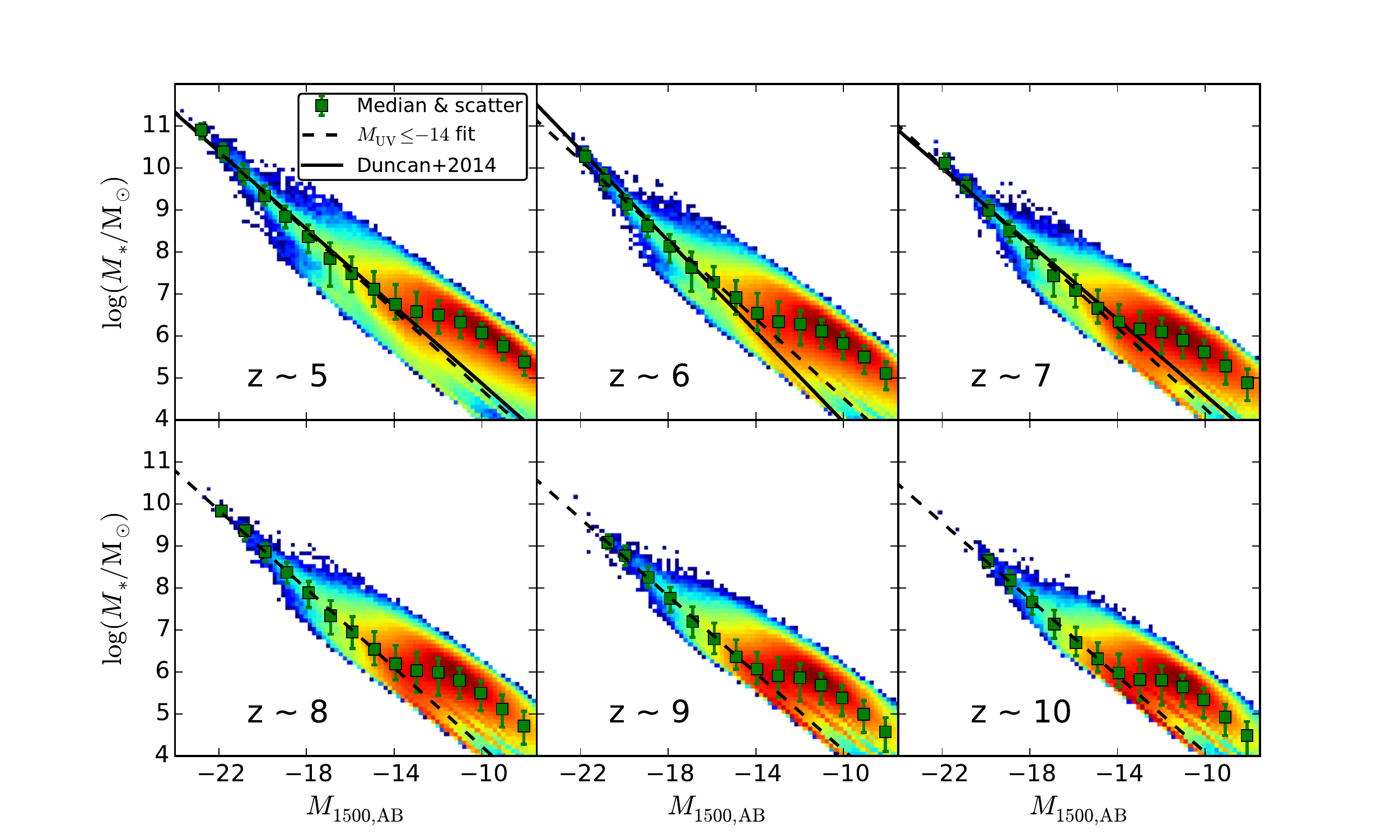}
    \caption{\label{fig:l-mg}
    Stellar masses of galaxies as a function of UV luminosity $M_{1500}$. The colour profile represents the logarithm density of the
    distribution. The green squares and error bars represent the median and 16th to 84th percentiles in bins
    which contain at least 5 model galaxies. The black dashed lines are the linear fit to the medians at $M_{1600}{\lesssim}{-14}$.
    The black lines show the observationally fit lines from \citet{2014MNRAS.444.2960D}, where the stellar mass is converted from a
    Chabrier IMF to Salpeter IMF by adding 0.24 to $\log(M_*)$. We see a close agreement between our model and observations
    at $z{\sim}5$--$7$ where the observational data is available.}
    \end{minipage}
\end{figure*}
Galaxies which continuously form stars naturally produce a relation between luminosity and stellar mass. Figure \ref{fig:l-mg}
shows stellar mass ($M_*$) as a function of observed UV luminosity ($M_{1500}$) for model galaxies predicted by \textsc{Meraxes}.
We see that our model predicts UV-bright galaxies to have large stellar masses. The model galaxies have stellar masses that are
distributed about the median at fixed luminosity with scatter ${\sim}0.2$--$0.5$ dex depending on UV luminosity. For comparison
we show the observed relation at $z{\sim}5$--$7$ from \citet{2014MNRAS.444.2960D}, who measured the stellar mass-luminosity
relation by fitting the observed photometric data with galaxy model SEDs. To compare with our model, the observational data
with nebular emission excluded from the SED fitting is used. We convert the observed data from a Chabrier IMF to a Salpeter
IMF by adding 0.24 to $\log (M_*)$ \citep{2014MNRAS.444.2960D}.
We linearly fit the $\log(M_*)$--$M_\mathrm{UV}$ relation for our bright galaxies with $M_{1500}{\leqslant}{-14}$ using
the relation\footnote{The $M_*$--$L_\mathrm{UV}$ relation can be derived by substituting
\mbox{$M_\mathrm{UV}={-2.5}\times\log_{10}(L_\mathrm{UV}\mathrm{[erg\,s^{-1}Hz^{-1}]})+51.6$}.}
\begin{equation}
\log{M_\mathrm{*}} = \frac{\mathrm{d}\log M_*}{\mathrm{d}M_\mathrm{UV}}
                               (M_\mathrm{UV} + 19.5) + \log M_{* (M_\mathrm{UV}{=}{-19.5})}.
\label{eqn:l-mg}
\end{equation}
\begin{table}
 \caption{The best-fit slopes and intercepts of the median $\log M_*$--$M_\mathrm{UV}$ relation (Equation \ref{eqn:l-mg}) for
          galaxies with $M_\mathrm{UV}{\leqslant}{-14}$.}\label{tab:l-mg}
 \centering
 \begin{tabular}{lcc}
  \hline
  \parbox{5mm}{$z$}& \parbox{28mm}{$\mathrm{d}\log M_*/\mathrm{d}M_\mathrm{UV}$}
                   & \parbox{28mm}{$\log M_{* (M_\mathrm{UV}{=}{-19.5})}$}\\
  \hline
       $5$ & $-0.474\pm 0.013$& $9.21\pm 0.04$\\
       $6$ & $-0.471\pm 0.021$& $9.00\pm 0.05$\\
       $7$ & $-0.477\pm 0.014$& $8.83\pm 0.04$\\
       $8$ & $-0.470\pm 0.010$& $8.68\pm 0.03$\\
       $9$ & $-0.459\pm 0.012$& $8.50\pm 0.04$\\
       $10$ & $-0.456\pm 0.013$& $8.41\pm 0.05$\\
  \hline
 \end{tabular}
\end{table}
\begin{figure}
    \includegraphics[width=\columnwidth]{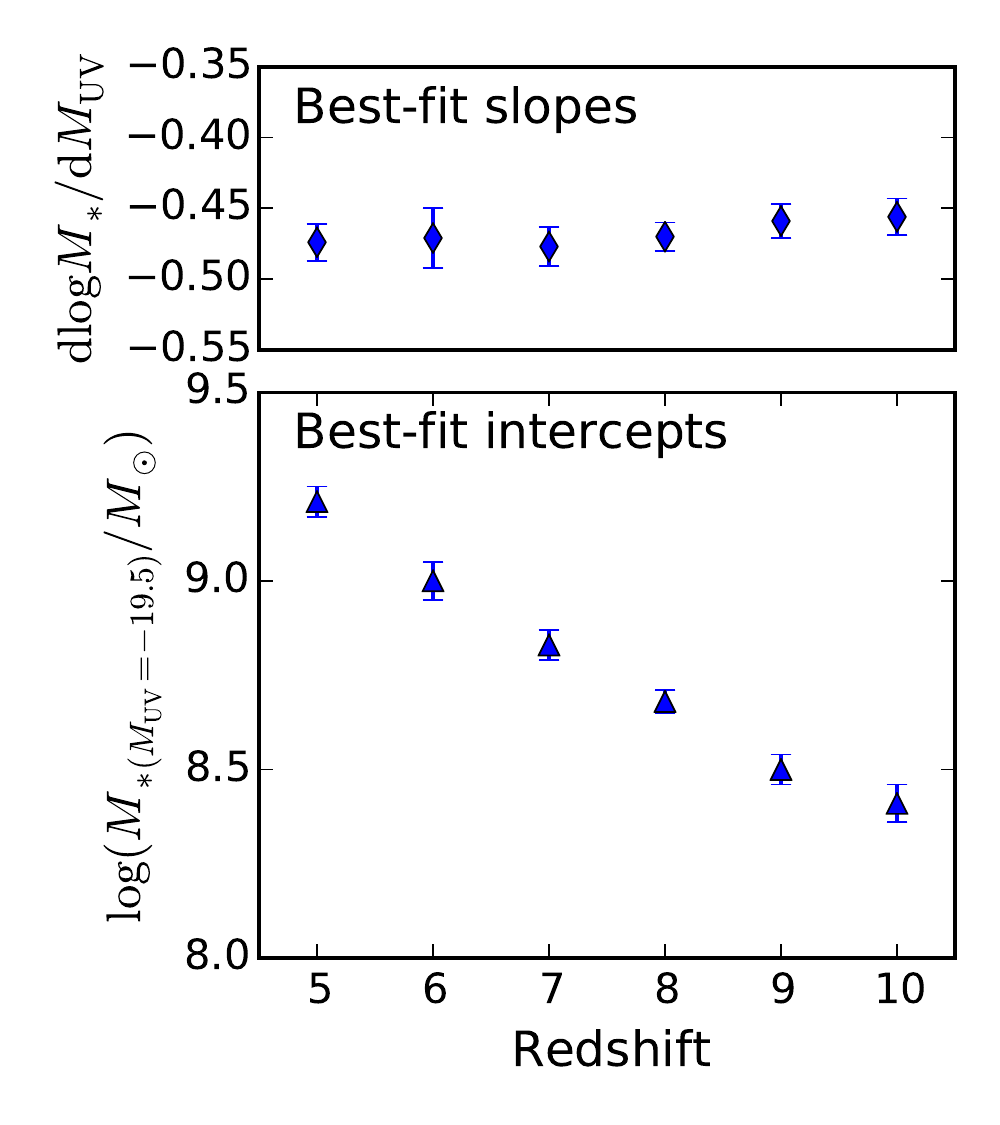}
    \caption{The best fitting slopes (upper panel) and the intercepts at $M_\mathrm{UV}{=}{-19.5}$ (lower panel) of the
             $\log M_*$--$M_\mathrm{UV}$ relation fit to the median trend for galaxies with $M_\mathrm{UV}{\leqslant}{-14}$.}
    \label{fig:mg_slope}
\end{figure}
The best-fit slopes and intercepts are shown in Table \ref{tab:l-mg} and Figure \ref{fig:mg_slope}.

At bright luminosities ($M_{1500}{\lesssim}{-14}$), the model galaxies are in good agreement with the observed mass--luminosity
relation. The best-fit slope (with median ${\sim}{-0.47}$) for model galaxies at $z{\sim}5$--$7$ does not significantly change with
redshift, and is close to the observed slopes of $\mathrm{d}\log M_*/\mathrm{d}M_\mathrm{UV}{\sim}-0.45$ to $-0.54$
from \citet{2014MNRAS.444.2960D} at $z{\sim}5$--$7$. Similar slopes are also found by
\citet{2012ApJ...752...66L} and \citet{2015ApJS..219...15S}.

We see that the best-fit intercept $\log M_{* (M_\mathrm{UV}{=}{-19.5})}$ evolves linearly with redshift. Assuming a constant slope
$\mathrm{d}\log M_*/\mathrm{d}M_\mathrm{UV}{=}{-0.47}$ for the bright galaxies at $z{\sim}5$--$10$,
we find the evolution of the luminosity--stellar mass relation can be estimated using the relation
\begin{equation}
\log\left(\frac{M_\mathrm{*}}{10^{8}\Msun}\right) = -0.47(M_{1500} + 19.5) - 0.15(z - 7) + 0.86.
\end{equation}
For fainter galaxies (${-14}{<}M_\mathrm{UV}{<}{-11}$) the slope of the luminosity--stellar mass relation changes
significantly to $\mathrm{d}\log M_*/\mathrm{d}M_\mathrm{UV}{\sim}{-0.1}$. This flattening of the
$\log M_*$--$M_\mathrm{UV}$ relation depends on the mass resolution of our dark matter N-body simulations.
The relation based on the higher resolution \emph{Tiny Tiamat} simulation flattens at $M_{1500}{>}{-13}$ (see Appendix \ref{app:a}).

\subsection{UV luminosity--halo mass relation}\label{sec:l-mvir}
\begin{figure*}
    \begin{minipage}{0.99\textwidth}
    \includegraphics[width=\columnwidth]{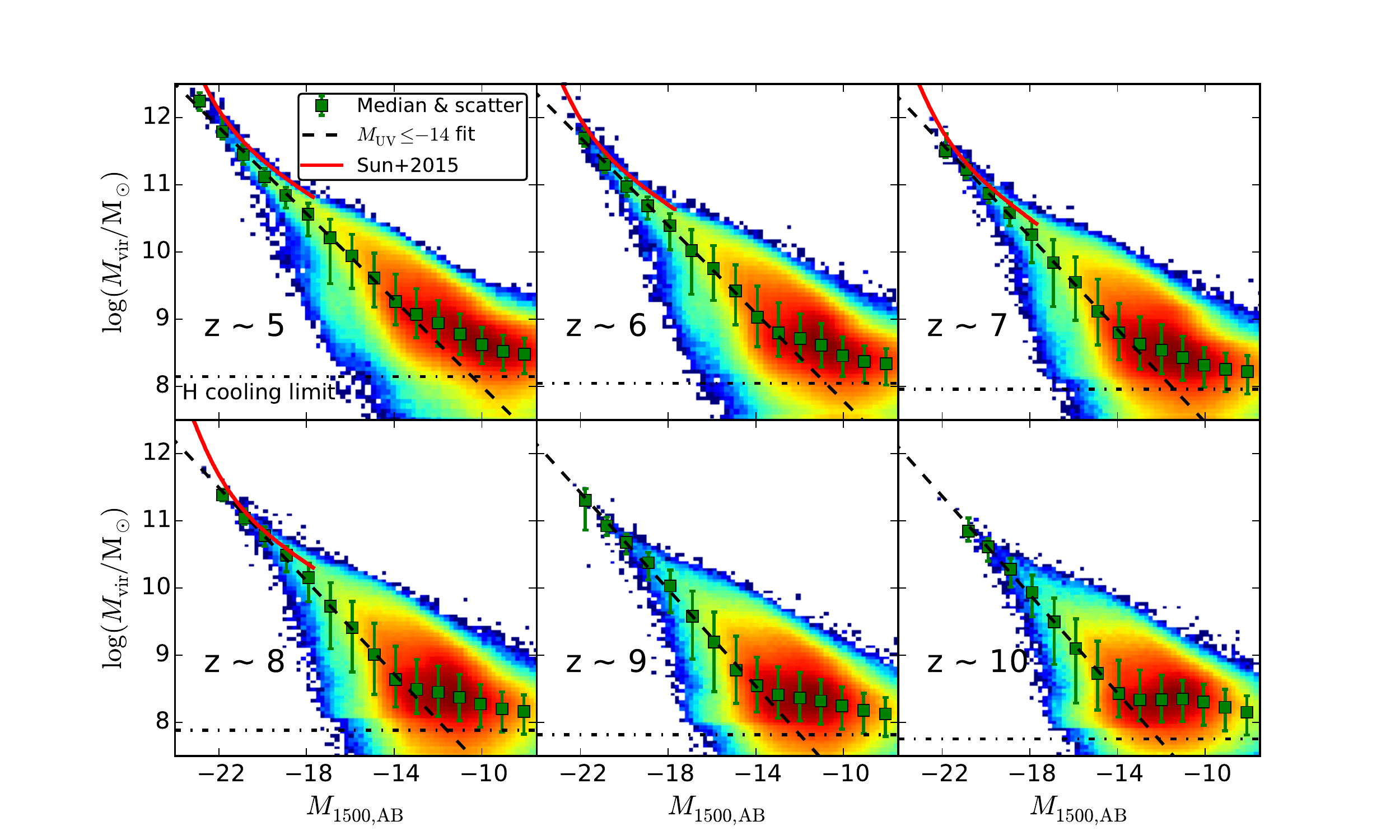}
    \caption{\label{fig:l-mvir}
    Dark matter halo (FoF group) masses as a function of UV luminosity. The colour profile represents the logarithm density
    of the distribution. The green squares and errorbars show the median and 16th to 84th percentiles in
    bins which contain at least 5 galaxies. The black dashed lines are the linear fit to the medians at $M_{1500}{\lesssim}{-14}$.
    The dash-dotted horizontal lines show the hydrogen cooling limit at each redshift. The red solid lines show the
    relation obtained using halo mass abundance matching technique from \citet{2015arXiv151206219S} at $z{\sim}5$--$8$.}
    \end{minipage}
\end{figure*}
\begin{figure}
    \includegraphics[width=\columnwidth]{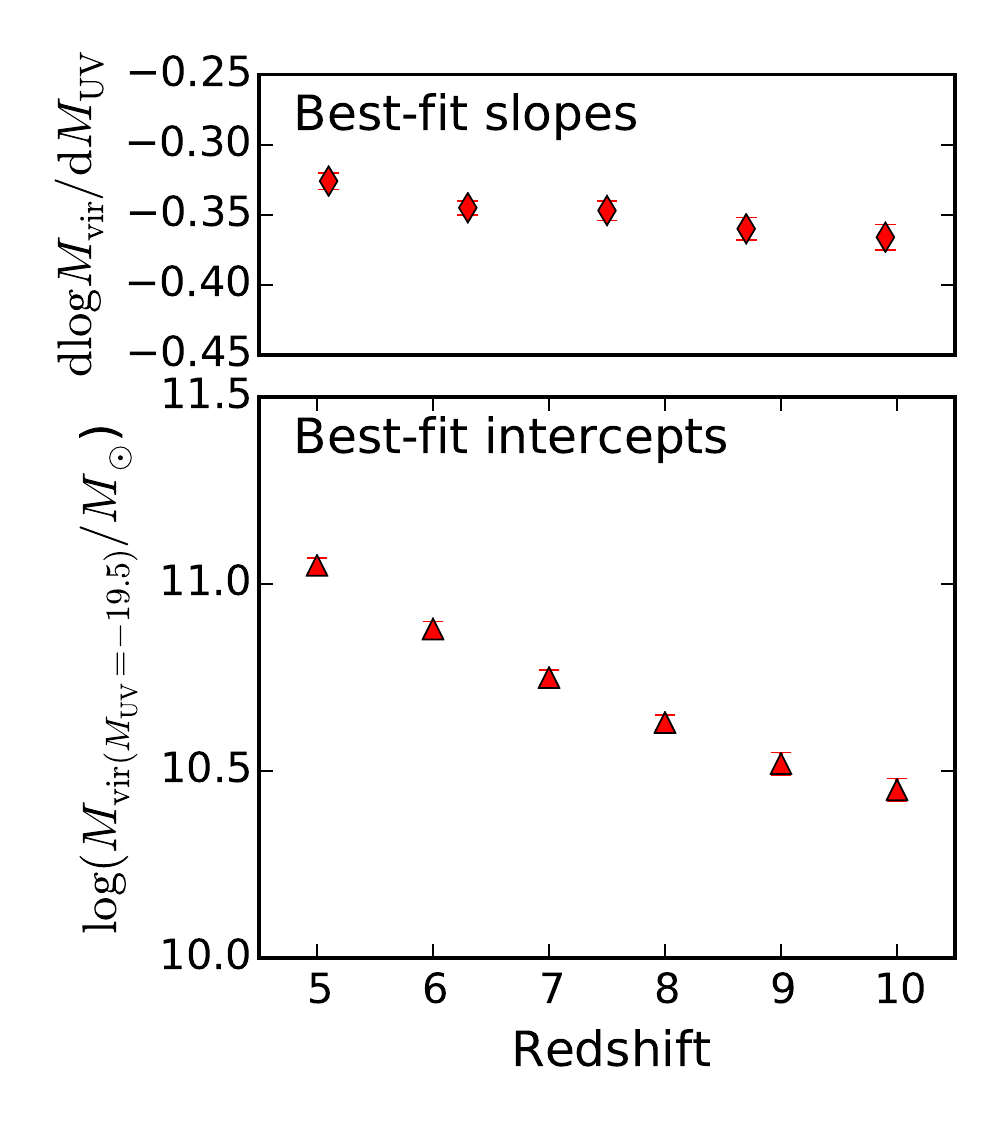}
    \caption{The best fitting slopes (upper panel) and the intercepts at $M_\mathrm{UV}{=}{-19.5}$ (lower panel) of
             the $\log M_\mathrm{vir}$--$M_\mathrm{UV}$ relation fit to the median trend for galaxies with $M_\mathrm{UV}{\leqslant}{-14}$.}
    \label{fig:mvir_slope}
\end{figure}

Before concluding, we discuss the relation between the masses of dark matter haloes (identified by the friends of friends
procedure, Paper-\Rom{1}) and the UV luminosity of hosted galaxies.
If a halo contains more than one galaxy, the luminosity is obtained by summing up all galaxies.
Figure \ref{fig:l-mvir} shows the halo mass-luminosity relation for all model galaxies at $z{\sim}5$--$10$. We see that
the UV-bright galaxies tend to be located in massive dark matter haloes. We fit the relation between $\log(M_\mathrm{vir})$ and
UV magnitude for galaxies brighter than ${-14}$ mag at each redshift using a linear relation, as shown by the dashed lines in
Figure \ref{fig:l-mvir}:
\begin{equation}
\log{M_\mathrm{vir}} = \frac{\mathrm{d}\log M_\mathrm{vir}}{\mathrm{d}M_\mathrm{UV}}
                               (M_\mathrm{UV} + 19.5) + \log M_{\mathrm{vir} (M_\mathrm{UV}{=}{-19.5})}
\label{eqn:l-mvir}
\end{equation}
The values of the best-fit slopes and intercepts are shown in Table \ref{tab:l-mvir} and Figure \ref{fig:mvir_slope}.
\begin{table}
 \caption{The best fitting slopes and the intercepts of the median $\log M_\mathrm{vir}$--$M_\mathrm{UV}$ relation
          (Equation \ref{eqn:l-mvir}) for galaxies with $M_\mathrm{UV}{\leqslant}{-14}$.}\label{tab:l-mvir}
 \centering
 \begin{tabular}{lcc}
  \hline
  \parbox{5mm}{$z$}& \parbox{28mm}{$\mathrm{d}\log M_\mathrm{vir}/\mathrm{d}M_\mathrm{UV}$}
                   & \parbox{28mm}{$\log M_{\mathrm{vir}, (M_\mathrm{UV}{=}{-19.5})}$}\\
  \hline
        $5$ & $-0.321\pm 0.006$& $11.05\pm 0.02$\\
        $6$ & $-0.326\pm 0.006$& $10.88\pm 0.02$\\
        $7$ & $-0.345\pm 0.005$& $10.75\pm 0.02$\\
        $8$ & $-0.347\pm 0.007$& $10.63\pm 0.02$\\
        $9$ & $-0.360\pm 0.008$& $10.52\pm 0.03$\\
        $10$ & $-0.366\pm 0.009$& $10.45\pm 0.03$\\
  \hline
 \end{tabular}
\end{table}

We find that the slope of the lines (with median ${\sim}{-0.35}$) slightly steepens towards higher redshift,
and the best-fit intercept $\log M_{\mathrm{vir} (M_\mathrm{UV}{=}{-19.5})}$ evolves linearly with redshift.
Assuming a constant slope $\mathrm{d}\log M_\mathrm{vir}/\mathrm{d}M_\mathrm{UV}{=}-0.35$ for the UV-bright galaxies
at $z{\sim}5$--$10$, we find the evolution of luminosity--halo mass relation can be estimated using
\begin{equation}
\log\left(\frac{M_\mathrm{vir}}{10^{10}\Msun}\right) = -0.35(M_{1500} + 19.5) - 0.13(z - 7) + 0.79.
\end{equation}
The scatter of the distribution at high luminosities is much smaller than at lower luminosities, varying from ${\sim 0.2}$ dex
at $M_{1500}{=}{-20}$ to ${\sim}0.5$ dex at $M_{1500}{=}{-16}$. Galaxies with the same UV
luminosity tend to be hosted by dark matter haloes with lower masses at higher redshift. This is partly because the stellar
populations of galaxies are generally younger at higher redshift. At $z{\sim}6$, galaxies with $M_{1500}{=}{-20}$
reside in haloes with a mass of ${\sim}10^{11.0\pm 0.1}\Msun$. A detailed analysis of clustering with \textsc{Meraxes} will be
presented in Park et al. (2015, in preparation). However, this
value is in agreement with the clustering analysis in \citet{2014ApJ...793...17B}. Figure \ref{fig:l-mvir} shows that at the
lowest luminosities, halo masses remain constant at ${\sim}10^8\Msun$. This is because the model prevents star formation in
haloes below the hydrogen-cooling limit, and explains why the LF turns over at low luminosities. This mass scale is well
resolved for the model based on the \emph{Tiny Tiamat} simulation (see Appendix \ref{app:a}). We plot the cooling limit for
dark matter haloes using the relation between $T_\mathrm{vir}$ and $M_\mathrm{vir}$ at high redshifts provided by \citet{2001PhR...349..125B}
and assuming $T_\mathrm{cool}{=}10^4$ K.

We note that the $\log M_\mathrm{vir}$--$M_\mathrm{UV}$ distribution is well described by a linear relation for ${-22}{<}M_\mathrm{UV}{<}{-14}$
and host halo masses of $M_\mathrm{vir}{<}10^{12}\Msun$. We do not have a significant sample of $M_\mathrm{vir}{>}10^{12}\Msun$
haloes, due to the simulation volume and so do not have reliable predictions in that mass range. Moreover, for haloes more massive
than $10^{12}\Msun$, AGN feedback, which is not included in the present simulations, may play a significant role in suppressing
star formation in galaxies and lead to a steep slope of the $\log M_\mathrm{vir}$--$M_\mathrm{UV}$ relation.

The $\log M_\mathrm{vir}$--$M_\mathrm{UV}$ relation can also be estimated from observations using the halo abundance
matching (HAM) technique \citep[e.g.][]{2004MNRAS.353..189V, 2015ApJ...813...21M, 2015arXiv151206219S, 2016MNRAS.455.2101M}.
We plot the $\log M_\mathrm{vir}$--$M_\mathrm{UV}$ relation from \citet{2015arXiv151206219S} at $z{\sim}5$--$8$ in
Figure \ref{fig:l-mvir} for comparison with our results. We see that the relations from our model agree with
HAM results within statistical uncertainty at $M_\mathrm{UV}{>}{-22}$. We do see a small difference at $M_\mathrm{UV}{>}{-19}$ between
these two models, which may have arisen from two systematic errors in the HAM methodology: (\rom{1}) assuming a monotonic
relation with an incorrect estimate of scatter between halo mass and galaxy luminosity; (\rom{2}) neglecting multiple halo
occupation, which leads to the sub-halo luminosities paired with central haloes.

\section{Conclusions}\label{sec:summary}
In this paper we have presented UV LFs for model galaxies during the EoR predicted by the semi-analytic model
\textsc{Meraxes}. \textsc{Meraxes} is a new semi-analytic model designed for studying reionization, and includes spatially-dependent
UVB feedback as well as time-dependent supernova feedback (Paper-\Rom{3}). By integrating model SEDs with the star formation
histories generated from our semi-analytic model, and including Ly$\alpha$ IGM absorption as well as a luminosity-dependent
dust attenuation, we derive UV luminosities, $M_{1600}$, for galaxies at $z{\sim}5$--$10$. We mimic the LBG colour selection
processes for our model galaxies, and obtain predictions for observed UV LFs. We also investigate relations
between the UV luminosity and a series of galaxy properties including the luminosity-SFR relation, luminosity-stellar mass
relation, and the luminosity-halo mass relation. We find that:
\begin{enumerate}
\item Having been calibrated to the stellar mass function at $z{\sim}5$--$7$, our model successfully reproduces the UV LF for
      high-redshift star-forming galaxies at $z{\sim}5$--$10$. The slope of predicted UV LFs remains steep below current detection
      limits, and becomes flat at $M_\mathrm{UV}{>}{-14}$ before declining below $M_\mathrm{UV}{\sim}{-12}$. This prediction will
      be testable in the future based on observations of faint galaxies with \textit{James Webb Space Telescope} and lensing.
\item The majority ($84$--$92$ per cent) of UV flux at $z{\sim}5$--$10$ is produced in galaxies with $M_\mathrm{UV}{<}{-13}$.
      At $z{\sim}5$, the flux is dominated by the UV-bright galaxies ($M_\mathrm{UV}{\leqslant}{-17}$).
      At $z{\geq}7$, galaxies with $-17{\lesssim}M_\mathrm{UV}{\lesssim}{-13}$ are the dominant contributors of UV flux.
\item Model galaxies with $M_\mathrm{UV}{\lesssim}{-14}$ are distributed around the luminosity--SFR relation
      from \citet{1998ARA&A..36..189K} and \citet{1998ApJ...498..106M} with a scatter of $0.1$--$0.3$dex. However, we find
      that the conversion between the high-redshift UV luminosity and SFR functions will be significantly biased by
      unaccounted-for scatter in the luminosity-SFR distribution for low-SFR galaxies (SFR${<}{0.01}\Msun\mathrm{yr}^{-1}$).
\item Model galaxies with $M_\mathrm{UV}{\leqslant}{-14}$ have stellar mass to luminosity relations that are consistent with
      the observed relations from \citet{2014MNRAS.444.2960D} at $z{\sim}5$--$7$. The $\log(M_*)$--$M_\mathrm{UV}$
      relation has a slope of ${\sim}{-0.47}$.
\item For high-mass dark matter haloes, there is a linear relation between $\log(M_\mathrm{vir})$ and $M_\mathrm{UV}$ with a slope of
      ${\sim}{-0.35}$. The scatter in this relation decreases with increasing luminosity. Galaxies with luminosities of $M_\mathrm{UV}{=}{-20}$
      at $z{\sim}6$ are hosted in dark matter haloes of mass $M_\mathrm{vir}{\sim}10^{11.0\pm 0.1}\Msun$. This mass decreases linearly
      towards higher redshift until $z{\sim}9$.
\end{enumerate}

In summary, the \textsc{Meraxes} semi-analytic model successfully describes the observed build up of the stellar mass of galaxies
during reionization as recorded in the redshift-dependent luminosity functions and stellar mass functions. This gives us
confidence in extrapolating the model to redshifts and luminosities beyond current observations. Future papers will apply
\textsc{Meraxes} to a series of additional observables during EoR including the morphology of 21-cm emission from the
IGM \citep[Paper-\Rom{5}]{2015arXiv151200564G} and galaxy sizes and clustering.

\section*{Acknowledgments}
This research was supported by the Victorian Life Sciences Computation Initiative (VLSCI), grant ref. UOM0005, on its Peak Computing Facility hosted at the University of Melbourne, an initiative of the Victorian Government, Australia. Part of this work was performed on the gSTAR national facility at Swinburne University of Technology. gSTAR is funded by Swinburne and the Australian Governments Education Investment Fund. This research program is funded by the Australian Research Council through the ARC Laureate Fellowship FL110100072 awarded to JSBW. AM acknowledges support from the European Research Council (ERC) under the European Unions Horizon 2020 research and innovation program (grant agreement No 638809 AIDA).

\bibliographystyle{mn2e}
\bibliography{reference/reference}

\begin{thebibliography}{93}
\expandafter\ifx\csname natexlab\endcsname\relax\def\natexlab#1{#1}\fi

\bibitem[{{Angel} {et~al}\mbox{.}(2015){Angel}, {Poole}, {Ludlow}, {Duffy},
  {Geil}, {Mutch}, {Mesinger}, \& {Wyithe}}]{2015arXiv151200560A}
{Angel} P.~W., {Poole} G.~B., {Ludlow} A.~D., {Duffy} A.~R., {Geil} P.~M.,
  {Mutch} S.~J., {Mesinger} A., {Wyithe} J.~S.~B., 2015, arXiv:1512.00560

\bibitem[{{Atek} {et~al}\mbox{.}(2015){Atek}, {Richard}, {Jauzac}, {Kneib},
  {Natarajan}, {Limousin}, {Schaerer}, {Jullo}, {Ebeling}, {Egami}, \&
  {Clement}}]{2015ApJ...814...69A}
{Atek} H. {et~al.}, 2015, \apj, 814, 69

\bibitem[{{Barkana} \& {Loeb}(2001)}]{2001PhR...349..125B}
{Barkana} R., {Loeb} A., 2001, \physrep, 349, 125

\bibitem[{{Barone-Nugent} {et~al}\mbox{.}(2014){Barone-Nugent}, {Trenti},
  {Wyithe}, {Bouwens}, {Oesch}, {Illingworth}, {Carollo}, {Su}, {Stiavelli},
  {Labbe}, \& {van Dokkum}}]{2014ApJ...793...17B}
{Barone-Nugent} R.~L. {et~al.}, 2014, \apj, 793, 17

\bibitem[{{Baugh}(2006)}]{2006RPPh...69.3101B}
{Baugh} C.~M., 2006, Reports on Progress in Physics, 69, 3101

\bibitem[{{Bouwens} {et~al}\mbox{.}(2014{\natexlab{a}}){Bouwens}, {Bradley},
  {Zitrin}, {Coe}, {Franx}, {Zheng}, {Smit}, {Host}, {Postman}, {Moustakas},
  {Labb{\'e}}, {Carrasco}, {Molino}, {Donahue}, {Kelson}, {Meneghetti},
  {Ben{\'{\i}}tez}, {Lemze}, {Umetsu}, {Broadhurst}, {Moustakas}, {Rosati},
  {Jouvel}, {Bartelmann}, {Ford}, {Graves}, {Grillo}, {Infante},
  {Jimenez-Teja}, {Lahav}, {Maoz}, {Medezinski}, {Melchior}, {Merten},
  {Nonino}, {Ogaz}, \& {Seitz}}]{2014ApJ...795..126B}
{Bouwens} R.~J. {et~al.}, 2014{\natexlab{a}}, \apj, 795, 126

\bibitem[{{Bouwens} {et~al}\mbox{.}(2007){Bouwens}, {Illingworth}, {Franx}, \&
  {Ford}}]{2007ApJ...670..928B}
{Bouwens} R.~J., {Illingworth} G.~D., {Franx} M., {Ford} H., 2007, \apj, 670,
  928

\bibitem[{{Bouwens} {et~al}\mbox{.}(2015{\natexlab{a}}){Bouwens},
  {Illingworth}, {Oesch}, {Caruana}, {Holwerda}, {Smit}, \&
  {Wilkins}}]{2015ApJ...811..140B}
{Bouwens} R.~J., {Illingworth} G.~D., {Oesch} P.~A., {Caruana} J., {Holwerda}
  B., {Smit} R., {Wilkins} S., 2015{\natexlab{a}}, \apj, 811, 140

\bibitem[{{Bouwens} {et~al}\mbox{.}(2012){Bouwens}, {Illingworth}, {Oesch},
  {Franx}, {Labb{\'e}}, {Trenti}, {van Dokkum}, {Carollo}, {Gonz{\'a}lez},
  {Smit}, \& {Magee}}]{2012ApJ...754...83B}
{Bouwens} R.~J. {et~al.}, 2012, \apj, 754, 83

\bibitem[{{Bouwens} {et~al}\mbox{.}(2011){Bouwens}, {Illingworth}, {Oesch},
  {Labb{\'e}}, {Trenti}, {van Dokkum}, {Franx}, {Stiavelli}, {Carollo},
  {Magee}, \& {Gonzalez}}]{2011ApJ...737...90B}
{Bouwens} R.~J. {et~al.}, 2011, \apj, 737, 90

\bibitem[{{Bouwens} {et~al}\mbox{.}(2014{\natexlab{b}}){Bouwens},
  {Illingworth}, {Oesch}, {Labb{\'e}}, {van Dokkum}, {Trenti}, {Franx}, {Smit},
  {Gonzalez}, \& {Magee}}]{2014ApJ...793..115B}
{Bouwens} R.~J. {et~al.}, 2014{\natexlab{b}}, \apj, 793, 115

\bibitem[{{Bouwens} {et~al}\mbox{.}(2010){Bouwens}, {Illingworth}, {Oesch},
  {Stiavelli}, {van Dokkum}, {Trenti}, {Magee}, {Labb{\'e}}, {Franx},
  {Carollo}, \& {Gonzalez}}]{2010ApJ...709L.133B}
{Bouwens} R.~J. {et~al.}, 2010, \apjl, 709, L133

\bibitem[{{Bouwens} {et~al}\mbox{.}(2015{\natexlab{b}}){Bouwens},
  {Illingworth}, {Oesch}, {Trenti}, {Labb{\'e}}, {Bradley}, {Carollo}, {van
  Dokkum}, {Gonzalez}, {Holwerda}, {Franx}, {Spitler}, {Smit}, \&
  {Magee}}]{2015ApJ...803...34B}
{Bouwens} R.~J. {et~al.}, 2015{\natexlab{b}}, \apj, 803, 34

\bibitem[{{Bouwens} {et~al}\mbox{.}(2015{\natexlab{c}}){Bouwens}, {Oesch},
  {Labbe}, {Illingworth}, {Fazio}, {Coe}, {Holwerda}, {Smit}, {Stefanon}, {van
  Dokkum}, {Trenti}, {Ashby}, {Huang}, {Spitler}, {Straatman}, {Bradley}, \&
  {Magee}}]{2015arXiv150601035B}
{Bouwens} R.~J. {et~al.}, 2015{\natexlab{c}}, arXiv:1506.01035

\bibitem[{{Bower} {et~al}\mbox{.}(2006){Bower}, {Benson}, {Malbon}, {Helly},
  {Frenk}, {Baugh}, {Cole}, \& {Lacey}}]{2006MNRAS.370..645B}
{Bower} R.~G., {Benson} A.~J., {Malbon} R., {Helly} J.~C., {Frenk} C.~S.,
  {Baugh} C.~M., {Cole} S., {Lacey} C.~G., 2006, \mnras, 370, 645

\bibitem[{{Bruzual} \& {Charlot}(2003)}]{2003MNRAS.344.1000B}
{Bruzual} G., {Charlot} S., 2003, \mnras, 344, 1000

\bibitem[{{Calzetti} {et~al}\mbox{.}(2000){Calzetti}, {Armus}, {Bohlin},
  {Kinney}, {Koornneef}, \& {Storchi-Bergmann}}]{2000ApJ...533..682C}
{Calzetti} D., {Armus} L., {Bohlin} R.~C., {Kinney} A.~L., {Koornneef} J.,
  {Storchi-Bergmann} T., 2000, \apj, 533, 682

\bibitem[{{Charlot} \& {Fall}(2000)}]{2000ApJ...539..718C}
{Charlot} S., {Fall} S.~M., 2000, \apj, 539, 718

\bibitem[{{Cole} {et~al}\mbox{.}(1994){Cole}, {Aragon-Salamanca}, {Frenk},
  {Navarro}, \& {Zepf}}]{1994MNRAS.271..781C}
{Cole} S., {Aragon-Salamanca} A., {Frenk} C.~S., {Navarro} J.~F., {Zepf} S.~E.,
  1994, \mnras, 271, 781

\bibitem[{{Cole} {et~al}\mbox{.}(2000){Cole}, {Lacey}, {Baugh}, \&
  {Frenk}}]{2000MNRAS.319..168C}
{Cole} S., {Lacey} C.~G., {Baugh} C.~M., {Frenk} C.~S., 2000, \mnras, 319, 168

\bibitem[{{Croton} {et~al}\mbox{.}(2006){Croton}, {Springel}, {White}, {De
  Lucia}, {Frenk}, {Gao}, {Jenkins}, {Kauffmann}, {Navarro}, \&
  {Yoshida}}]{2006MNRAS.365...11C}
{Croton} D.~J. {et~al.}, 2006, \mnras, 365, 11

\bibitem[{{Duffy} {et~al}\mbox{.}(2014){Duffy}, {Wyithe}, {Mutch}, \&
  {Poole}}]{2014MNRAS.443.3435D}
{Duffy} A.~R., {Wyithe} J.~S.~B., {Mutch} S.~J., {Poole} G.~B., 2014, \mnras,
  443, 3435

\bibitem[{{Duncan} {et~al}\mbox{.}(2014){Duncan}, {Conselice}, {Mortlock},
  {Hartley}, {Guo}, {Ferguson}, {Dav{\'e}}, {Lu}, {Ownsworth}, {Ashby},
  {Dekel}, {Dickinson}, {Faber}, {Giavalisco}, {Grogin}, {Kocevski},
  {Koekemoer}, {Somerville}, \& {White}}]{2014MNRAS.444.2960D}
{Duncan} K. {et~al.}, 2014, \mnras, 444, 2960

\bibitem[{{Fall} \& {Efstathiou}(1980)}]{1980MNRAS.193..189F}
{Fall} S.~M., {Efstathiou} G., 1980, \mnras, 193, 189

\bibitem[{{Fan} {et~al}\mbox{.}(2006{\natexlab{a}}){Fan}, {Carilli}, \&
  {Keating}}]{2006ARA&A..44..415F}
{Fan} X., {Carilli} C.~L., {Keating} B., 2006{\natexlab{a}}, \araa, 44, 415

\bibitem[{{Fan} {et~al}\mbox{.}(2006{\natexlab{b}}){Fan}, {Strauss}, {Becker},
  {White}, {Gunn}, {Knapp}, {Richards}, {Schneider}, {Brinkmann}, \&
  {Fukugita}}]{2006AJ....132..117F}
{Fan} X. {et~al.}, 2006{\natexlab{b}}, \aj, 132, 117

\bibitem[{{Finkelstein} {et~al}\mbox{.}(2012){Finkelstein}, {Papovich},
  {Salmon}, {Finlator}, {Dickinson}, {Ferguson}, {Giavalisco}, {Koekemoer},
  {Reddy}, {Bassett}, {Conselice}, {Dunlop}, {Faber}, {Grogin}, {Hathi},
  {Kocevski}, {Lai}, {Lee}, {McLure}, {Mobasher}, \&
  {Newman}}]{2012ApJ...756..164F}
{Finkelstein} S.~L. {et~al.}, 2012, \apj, 756, 164

\bibitem[{{Geil} {et~al}\mbox{.}(2015){Geil}, {Mutch}, {Poole}, {Angel},
  {Duffy}, {Mesinger}, \& {Wyithe}}]{2015arXiv151200564G}
{Geil} P.~M., {Mutch} S.~J., {Poole} G.~B., {Angel} P.~W., {Duffy} A.~R.,
  {Mesinger} A., {Wyithe} J.~S.~B., 2015, arXiv:1512.00564

\bibitem[{{Gonz{\'a}lez} {et~al}\mbox{.}(2011){Gonz{\'a}lez}, {Labb{\'e}},
  {Bouwens}, {Illingworth}, {Franx}, \& {Kriek}}]{2011ApJ...735L..34G}
{Gonz{\'a}lez} V., {Labb{\'e}} I., {Bouwens} R.~J., {Illingworth} G., {Franx}
  M., {Kriek} M., 2011, \apjl, 735, L34

\bibitem[{{Gonzalez-Perez} {et~al}\mbox{.}(2014){Gonzalez-Perez}, {Lacey},
  {Baugh}, {Lagos}, {Helly}, {Campbell}, \& {Mitchell}}]{2014MNRAS.439..264G}
{Gonzalez-Perez} V., {Lacey} C.~G., {Baugh} C.~M., {Lagos} C.~D.~P., {Helly}
  J., {Campbell} D.~J.~R., {Mitchell} P.~D., 2014, \mnras, 439, 264

\bibitem[{{Grazian} {et~al}\mbox{.}(2015){Grazian}, {Fontana}, {Santini},
  {Dunlop}, {Ferguson}, {Castellano}, {Amorin}, {Ashby}, {Barro}, {Behroozi},
  {Boutsia}, {Caputi}, {Chary}, {Dekel}, {Dickinson}, {Faber}, {Fazio},
  {Finkelstein}, {Galametz}, {Giallongo}, {Giavalisco}, {Grogin}, {Guo},
  {Kocevski}, {Koekemoer}, {Koo}, {Lee}, {Lu}, {Merlin}, {Mobasher}, {Nonino},
  {Papovich}, {Paris}, {Pentericci}, {Reddy}, {Renzini}, {Salmon}, {Salvato},
  {Sommariva}, {Song}, \& {Vanzella}}]{2015A&A...575A..96G}
{Grazian} A. {et~al.}, 2015, \aap, 575, A96

\bibitem[{{Heinis} {et~al}\mbox{.}(2014){Heinis}, {Buat}, {B{\'e}thermin},
  {Bock}, {Burgarella}, {Conley}, {Cooray}, {Farrah}, {Ilbert}, {Magdis},
  {Marsden}, {Oliver}, {Rigopoulou}, {Roehlly}, {Schulz}, {Symeonidis},
  {Viero}, {Xu}, \& {Zemcov}}]{2014MNRAS.437.1268H}
{Heinis} S. {et~al.}, 2014, \mnras, 437, 1268

\bibitem[{{Kauffmann} {et~al}\mbox{.}(1999){Kauffmann}, {Colberg}, {Diaferio},
  \& {White}}]{1999MNRAS.303..188K}
{Kauffmann} G., {Colberg} J.~M., {Diaferio} A., {White} S.~D.~M., 1999, \mnras,
  303, 188

\bibitem[{{Kauffmann} {et~al}\mbox{.}(1993){Kauffmann}, {White}, \&
  {Guiderdoni}}]{1993MNRAS.264..201K}
{Kauffmann} G., {White} S.~D.~M., {Guiderdoni} B., 1993, \mnras, 264, 201

\bibitem[{{Kennicutt}(1998)}]{1998ARA&A..36..189K}
{Kennicutt}, Jr. R.~C., 1998, \araa, 36, 189

\bibitem[{{Kuhlen} \& {Faucher-Gigu{\`e}re}(2012)}]{2012MNRAS.423..862K}
{Kuhlen} M., {Faucher-Gigu{\`e}re} C.-A., 2012, \mnras, 423, 862

\bibitem[{{Lacey} {et~al}\mbox{.}(2011){Lacey}, {Baugh}, {Frenk}, \&
  {Benson}}]{2011MNRAS.412.1828L}
{Lacey} C.~G., {Baugh} C.~M., {Frenk} C.~S., {Benson} A.~J., 2011, \mnras, 412,
  1828

\bibitem[{{Lacey} {et~al}\mbox{.}(2015){Lacey}, {Baugh}, {Frenk}, {Benson},
  {Bower}, {Cole}, {Gonzalez-Perez}, {Helly}, {Lagos}, \&
  {Mitchell}}]{2015arXiv150908473L}
{Lacey} C.~G. {et~al.}, 2015, arXiv:1509.08473

\bibitem[{{Lee} {et~al}\mbox{.}(2012){Lee}, {Ferguson}, {Wiklind}, {Dahlen},
  {Dickinson}, {Giavalisco}, {Grogin}, {Papovich}, {Messias}, {Guo}, \&
  {Lin}}]{2012ApJ...752...66L}
{Lee} K.-S. {et~al.}, 2012, \apj, 752, 66

\bibitem[{{Leitherer} {et~al}\mbox{.}(2014){Leitherer}, {Ekstr{\"o}m},
  {Meynet}, {Schaerer}, {Agienko}, \& {Levesque}}]{2014ApJS..212...14L}
{Leitherer} C., {Ekstr{\"o}m} S., {Meynet} G., {Schaerer} D., {Agienko} K.~B.,
  {Levesque} E.~M., 2014, \apjs, 212, 14

\bibitem[{{Leitherer} \& {Heckman}(1995)}]{1995ApJS...96....9L}
{Leitherer} C., {Heckman} T.~M., 1995, \apjs, 96, 9

\bibitem[{{Leitherer} {et~al}\mbox{.}(2010){Leitherer}, {Ortiz Ot{\'a}lvaro},
  {Bresolin}, {Kudritzki}, {Lo Faro}, {Pauldrach}, {Pettini}, \&
  {Rix}}]{2010ApJS..189..309L}
{Leitherer} C., {Ortiz Ot{\'a}lvaro} P.~A., {Bresolin} F., {Kudritzki} R.-P.,
  {Lo Faro} B., {Pauldrach} A.~W.~A., {Pettini} M., {Rix} S.~A., 2010, \apjs,
  189, 309

\bibitem[{{Leitherer} {et~al}\mbox{.}(1999){Leitherer}, {Schaerer}, {Goldader},
  {Delgado}, {Robert}, {Kune}, {de Mello}, {Devost}, \&
  {Heckman}}]{1999ApJS..123....3L}
{Leitherer} C. {et~al.}, 1999, \apjs, 123, 3

\bibitem[{{Madau} {et~al}\mbox{.}(1998){Madau}, {Pozzetti}, \&
  {Dickinson}}]{1998ApJ...498..106M}
{Madau} P., {Pozzetti} L., {Dickinson} M., 1998, \apj, 498, 106

\bibitem[{{Mashian} {et~al}\mbox{.}(2016){Mashian}, {Oesch}, \&
  {Loeb}}]{2016MNRAS.455.2101M}
{Mashian} N., {Oesch} P.~A., {Loeb} A., 2016, \mnras, 455, 2101

\bibitem[{{Mason} {et~al}\mbox{.}(2015){Mason}, {Trenti}, \&
  {Treu}}]{2015ApJ...813...21M}
{Mason} C.~A., {Trenti} M., {Treu} T., 2015, \apj, 813, 21

\bibitem[{{McLure} {et~al}\mbox{.}(2013){McLure}, {Dunlop}, {Bowler},
  {Curtis-Lake}, {Schenker}, {Ellis}, {Robertson}, {Koekemoer}, {Rogers},
  {Ono}, {Ouchi}, {Charlot}, {Wild}, {Stark}, {Furlanetto}, {Cirasuolo}, \&
  {Targett}}]{2013MNRAS.432.2696M}
{McLure} R.~J. {et~al.}, 2013, \mnras, 432, 2696

\bibitem[{{McLure} {et~al}\mbox{.}(2010){McLure}, {Dunlop}, {Cirasuolo},
  {Koekemoer}, {Sabbi}, {Stark}, {Targett}, \& {Ellis}}]{2010MNRAS.403..960M}
{McLure} R.~J., {Dunlop} J.~S., {Cirasuolo} M., {Koekemoer} A.~M., {Sabbi} E.,
  {Stark} D.~P., {Targett} T.~A., {Ellis} R.~S., 2010, \mnras, 403, 960

\bibitem[{{Mesinger} \& {Furlanetto}(2007)}]{2007ApJ...669..663M}
{Mesinger} A., {Furlanetto} S., 2007, \apj, 669, 663

\bibitem[{{Mesinger} {et~al}\mbox{.}(2011){Mesinger}, {Furlanetto}, \&
  {Cen}}]{2011MNRAS.411..955M}
{Mesinger} A., {Furlanetto} S., {Cen} R., 2011, \mnras, 411, 955

\bibitem[{{Meurer} {et~al}\mbox{.}(1999){Meurer}, {Heckman}, \&
  {Calzetti}}]{1999ApJ...521...64M}
{Meurer} G.~R., {Heckman} T.~M., {Calzetti} D., 1999, \apj, 521, 64

\bibitem[{{Mo} {et~al}\mbox{.}(1998){Mo}, {Mao}, \&
  {White}}]{1998MNRAS.295..319M}
{Mo} H.~J., {Mao} S., {White} S.~D.~M., 1998, \mnras, 295, 319

\bibitem[{{Mu{\~n}oz} \& {Loeb}(2011)}]{2011ApJ...729...99M}
{Mu{\~n}oz} J.~A., {Loeb} A., 2011, \apj, 729, 99

\bibitem[{{Mutch} {et~al}\mbox{.}(2015){Mutch}, {Geil}, {Poole}, {Angel},
  {Duffy}, {Mesinger}, \& {Wyithe}}]{2015arXiv151200562M}
{Mutch} S.~J., {Geil} P.~M., {Poole} G.~B., {Angel} P.~W., {Duffy} A.~R.,
  {Mesinger} A., {Wyithe} J.~S.~B., 2015, arXiv:1512.00562

\bibitem[{{Mutch} {et~al}\mbox{.}(2013){Mutch}, {Poole}, \&
  {Croton}}]{2013MNRAS.428.2001M}
{Mutch} S.~J., {Poole} G.~B., {Croton} D.~J., 2013, \mnras, 428, 2001

\bibitem[{{Oesch} {et~al}\mbox{.}(2010){Oesch}, {Bouwens}, {Illingworth},
  {Carollo}, {Franx}, {Labb{\'e}}, {Magee}, {Stiavelli}, {Trenti}, \& {van
  Dokkum}}]{2010ApJ...709L..16O}
{Oesch} P.~A. {et~al.}, 2010, \apjl, 709, L16

\bibitem[{{Oke} \& {Gunn}(1983)}]{1983ApJ...266..713O}
{Oke} J.~B., {Gunn} J.~E., 1983, \apj, 266, 713

\bibitem[{{O'Shea} {et~al}\mbox{.}(2015){O'Shea}, {Wise}, {Xu}, \&
  {Norman}}]{2015ApJ...807L..12O}
{O'Shea} B.~W., {Wise} J.~H., {Xu} H., {Norman} M.~L., 2015, \apjl, 807, L12

\bibitem[{{Paardekooper} {et~al}\mbox{.}(2015){Paardekooper}, {Khochfar}, \&
  {Dalla Vecchia}}]{2015MNRAS.451.2544P}
{Paardekooper} J.-P., {Khochfar} S., {Dalla Vecchia} C., 2015, \mnras, 451,
  2544

\bibitem[{{Pannella} {et~al}\mbox{.}(2009){Pannella}, {Carilli}, {Daddi},
  {McCracken}, {Owen}, {Renzini}, {Strazzullo}, {Civano}, {Koekemoer},
  {Schinnerer}, {Scoville}, {Smol{\v c}i{\'c}}, {Taniguchi}, {Aussel}, {Kneib},
  {Ilbert}, {Mellier}, {Salvato}, {Thompson}, \&
  {Willott}}]{2009ApJ...698L.116P}
{Pannella} M. {et~al.}, 2009, \apjl, 698, L116

\bibitem[{{Planck Collaboration} {et~al}\mbox{.}(2015){Planck Collaboration},
  {Ade}, {Aghanim}, {Arnaud}, {Ashdown}, {Aumont}, {Baccigalupi}, {Banday},
  {Barreiro}, {Bartlett}, \& et~al.}]{2015arXiv150201589P}
{Planck Collaboration} {et~al.}, 2015, arXiv:1502.01589

\bibitem[{{Poole} {et~al}\mbox{.}(2015){Poole}, {Angel}, {Mutch}, {Power},
  {Duffy}, {Geil}, {Mesinger}, \& {Wyithe}}]{2015arXiv151200559P}
{Poole} G.~B., {Angel} P.~W., {Mutch} S.~J., {Power} C., {Duffy} A.~R., {Geil}
  P.~M., {Mesinger} A., {Wyithe} S.~B., 2015, arXiv:1512.00559

\bibitem[{{Reddy} {et~al}\mbox{.}(2006){Reddy}, {Steidel}, {Fadda}, {Yan},
  {Pettini}, {Shapley}, {Erb}, \& {Adelberger}}]{2006ApJ...644..792R}
{Reddy} N.~A., {Steidel} C.~C., {Fadda} D., {Yan} L., {Pettini} M., {Shapley}
  A.~E., {Erb} D.~K., {Adelberger} K.~L., 2006, \apj, 644, 792

\bibitem[{{Robertson} {et~al}\mbox{.}(2010){Robertson}, {Ellis}, {Dunlop},
  {McLure}, \& {Stark}}]{2010Natur.468...49R}
{Robertson} B.~E., {Ellis} R.~S., {Dunlop} J.~S., {McLure} R.~J., {Stark}
  D.~P., 2010, \nat, 468, 49

\bibitem[{{Robertson} {et~al}\mbox{.}(2015){Robertson}, {Ellis}, {Furlanetto},
  \& {Dunlop}}]{2015ApJ...802L..19R}
{Robertson} B.~E., {Ellis} R.~S., {Furlanetto} S.~R., {Dunlop} J.~S., 2015,
  \apjl, 802, L19

\bibitem[{{Robertson} {et~al}\mbox{.}(2013){Robertson}, {Furlanetto},
  {Schneider}, {Charlot}, {Ellis}, {Stark}, {McLure}, {Dunlop}, {Koekemoer},
  {Schenker}, {Ouchi}, {Ono}, {Curtis-Lake}, {Rogers}, {Bowler}, \&
  {Cirasuolo}}]{2013ApJ...768...71R}
{Robertson} B.~E. {et~al.}, 2013, \apj, 768, 71

\bibitem[{{Salpeter}(1955)}]{1955ApJ...121..161S}
{Salpeter} E.~E., 1955, \apj, 121, 161

\bibitem[{{Schaye} {et~al}\mbox{.}(2015){Schaye}, {Crain}, {Bower}, {Furlong},
  {Schaller}, {Theuns}, {Dalla Vecchia}, {Frenk}, {McCarthy}, {Helly},
  {Jenkins}, {Rosas-Guevara}, {White}, {Baes}, {Booth}, {Camps}, {Navarro},
  {Qu}, {Rahmati}, {Sawala}, {Thomas}, \& {Trayford}}]{2015MNRAS.446..521S}
{Schaye} J. {et~al.}, 2015, \mnras, 446, 521

\bibitem[{{Schaye} {et~al}\mbox{.}(2010){Schaye}, {Dalla Vecchia}, {Booth},
  {Wiersma}, {Theuns}, {Haas}, {Bertone}, {Duffy}, {McCarthy}, \& {van de
  Voort}}]{2010MNRAS.402.1536S}
{Schaye} J. {et~al.}, 2010, \mnras, 402, 1536

\bibitem[{{Schenker} {et~al}\mbox{.}(2013){Schenker}, {Robertson}, {Ellis},
  {Ono}, {McLure}, {Dunlop}, {Koekemoer}, {Bowler}, {Ouchi}, {Curtis-Lake},
  {Rogers}, {Schneider}, {Charlot}, {Stark}, {Furlanetto}, \&
  {Cirasuolo}}]{2013ApJ...768..196S}
{Schenker} M.~A. {et~al.}, 2013, \apj, 768, 196

\bibitem[{{Schmidt} {et~al}\mbox{.}(2014){Schmidt}, {Treu}, {Trenti},
  {Bradley}, {Kelly}, {Oesch}, {Holwerda}, {Shull}, \&
  {Stiavelli}}]{2014ApJ...786...57S}
{Schmidt} K.~B. {et~al.}, 2014, \apj, 786, 57

\bibitem[{{Shibuya} {et~al}\mbox{.}(2015){Shibuya}, {Ouchi}, \&
  {Harikane}}]{2015ApJS..219...15S}
{Shibuya} T., {Ouchi} M., {Harikane} Y., 2015, \apjs, 219, 15

\bibitem[{{Smit} {et~al}\mbox{.}(2012){Smit}, {Bouwens}, {Franx},
  {Illingworth}, {Labb{\'e}}, {Oesch}, \& {van Dokkum}}]{2012ApJ...756...14S}
{Smit} R., {Bouwens} R.~J., {Franx} M., {Illingworth} G.~D., {Labb{\'e}} I.,
  {Oesch} P.~A., {van Dokkum} P.~G., 2012, \apj, 756, 14

\bibitem[{{Sobacchi} \& {Mesinger}(2013{\natexlab{a}})}]{2013MNRAS.432.3340S}
{Sobacchi} E., {Mesinger} A., 2013{\natexlab{a}}, \mnras, 432, 3340

\bibitem[{{Sobacchi} \& {Mesinger}(2013{\natexlab{b}})}]{2013MNRAS.432L..51S}
{Sobacchi} E., {Mesinger} A., 2013{\natexlab{b}}, \mnras, 432, L51

\bibitem[{{Somerville} {et~al}\mbox{.}(2008){Somerville}, {Hopkins}, {Cox},
  {Robertson}, \& {Hernquist}}]{2008MNRAS.391..481S}
{Somerville} R.~S., {Hopkins} P.~F., {Cox} T.~J., {Robertson} B.~E.,
  {Hernquist} L., 2008, \mnras, 391, 481

\bibitem[{{Somerville} {et~al}\mbox{.}(2001){Somerville}, {Primack}, \&
  {Faber}}]{2001MNRAS.320..504S}
{Somerville} R.~S., {Primack} J.~R., {Faber} S.~M., 2001, \mnras, 320, 504

\bibitem[{{Song} {et~al}\mbox{.}(2015){Song}, {Finkelstein}, {Ashby},
  {Grazian}, {Lu}, {Papovich}, {Salmon}, {Somerville}, {Dickinson}, {Duncan},
  {Faber}, {Fazio}, {Ferguson}, {Fontana}, {Guo}, {Hathi}, {Lee}, {Merlin}, \&
  {Willner}}]{2015arXiv150705636S}
{Song} M. {et~al.}, 2015, arXiv:1507.05636

\bibitem[{{Springel} {et~al}\mbox{.}(2005){Springel}, {White}, {Jenkins},
  {Frenk}, {Yoshida}, {Gao}, {Navarro}, {Thacker}, {Croton}, {Helly},
  {Peacock}, {Cole}, {Thomas}, {Couchman}, {Evrard}, {Colberg}, \&
  {Pearce}}]{2005Natur.435..629S}
{Springel} V. {et~al.}, 2005, \nat, 435, 629

\bibitem[{{Steidel} {et~al}\mbox{.}(1999){Steidel}, {Adelberger}, {Giavalisco},
  {Dickinson}, \& {Pettini}}]{1999ApJ...519....1S}
{Steidel} C.~C., {Adelberger} K.~L., {Giavalisco} M., {Dickinson} M., {Pettini}
  M., 1999, \apj, 519, 1

\bibitem[{{Steidel} {et~al}\mbox{.}(1996){Steidel}, {Giavalisco}, {Pettini},
  {Dickinson}, \& {Adelberger}}]{1996ApJ...462L..17S}
{Steidel} C.~C., {Giavalisco} M., {Pettini} M., {Dickinson} M., {Adelberger}
  K.~L., 1996, \apjl, 462, L17

\bibitem[{{Sun} \& {Furlanetto}(2015)}]{2015arXiv151206219S}
{Sun} G., {Furlanetto} S.~R., 2015, arXiv:1512.06219

\bibitem[{{Sutherland} \& {Dopita}(1993)}]{1993ApJS...88..253S}
{Sutherland} R.~S., {Dopita} M.~A., 1993, \apjs, 88, 253

\bibitem[{{Vale} \& {Ostriker}(2004)}]{2004MNRAS.353..189V}
{Vale} A., {Ostriker} J.~P., 2004, \mnras, 353, 189

\bibitem[{{V{\'a}zquez} \& {Leitherer}(2005)}]{2005ApJ...621..695V}
{V{\'a}zquez} G.~A., {Leitherer} C., 2005, \apj, 621, 695

\bibitem[{{Weisz} {et~al}\mbox{.}(2014){Weisz}, {Johnson}, \&
  {Conroy}}]{2014ApJ...794L...3W}
{Weisz} D.~R., {Johnson} B.~D., {Conroy} C., 2014, \apjl, 794, L3

\bibitem[{{White} \& {Frenk}(1991)}]{1991ApJ...379...52W}
{White} S.~D.~M., {Frenk} C.~S., 1991, \apj, 379, 52

\bibitem[{{White} \& {Rees}(1978)}]{1978MNRAS.183..341W}
{White} S.~D.~M., {Rees} M.~J., 1978, \mnras, 183, 341

\bibitem[{{Wilkins} {et~al}\mbox{.}(2016){Wilkins}, {Bouwens}, {Oesch},
  {Labb{\'e}}, {Sargent}, {Caruana}, {Wardlow}, \&
  {Clay}}]{2016MNRAS.455..659W}
{Wilkins} S.~M., {Bouwens} R.~J., {Oesch} P.~A., {Labb{\'e}} I., {Sargent} M.,
  {Caruana} J., {Wardlow} J., {Clay} S., 2016, \mnras, 455, 659

\bibitem[{{Wilkins} {et~al}\mbox{.}(2011){Wilkins}, {Bunker}, {Lorenzoni}, \&
  {Caruana}}]{2011MNRAS.411...23W}
{Wilkins} S.~M., {Bunker} A.~J., {Lorenzoni} S., {Caruana} J., 2011, \mnras,
  411, 23

\bibitem[{{Wise} \& {Cen}(2009)}]{2009ApJ...693..984W}
{Wise} J.~H., {Cen} R., 2009, \apj, 693, 984

\bibitem[{{Wise} {et~al}\mbox{.}(2014){Wise}, {Demchenko}, {Halicek}, {Norman},
  {Turk}, {Abel}, \& {Smith}}]{2014MNRAS.442.2560W}
{Wise} J.~H., {Demchenko} V.~G., {Halicek} M.~T., {Norman} M.~L., {Turk} M.~J.,
  {Abel} T., {Smith} B.~D., 2014, \mnras, 442, 2560

\bibitem[{{Wise} {et~al}\mbox{.}(2012){Wise}, {Turk}, {Norman}, \&
  {Abel}}]{2012ApJ...745...50W}
{Wise} J.~H., {Turk} M.~J., {Norman} M.~L., {Abel} T., 2012, \apj, 745, 50

\end{thebibliography}

\appendix
\section{luminosity---Mass Relations based on \emph{Tiny Tiamat}}\label{app:a}
\begin{figure*}
    \begin{minipage}{0.99\textwidth}
    \includegraphics[width=\columnwidth]{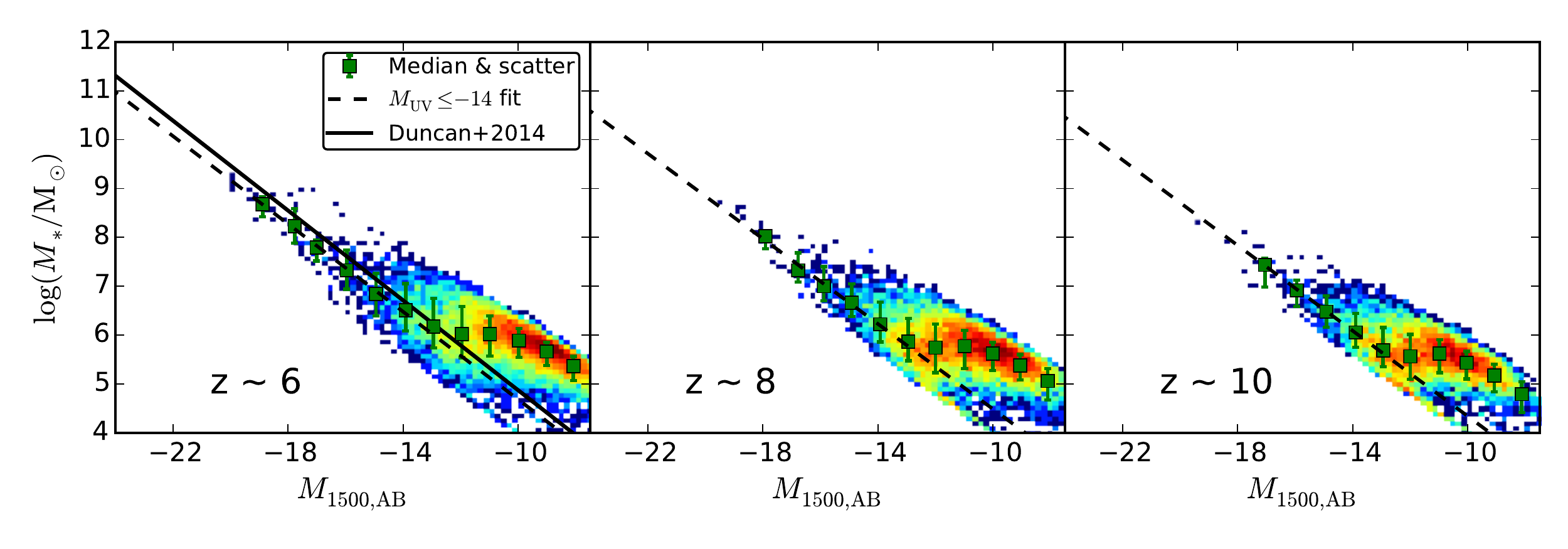}
    \caption{\label{fig:l-mg_tiny}
    Same as Figure \ref{fig:l-mg} but for model galaxies at $z{\sim}6$, $8$ and $10$ based on the \emph{Tiny Tiamat} N-body simulation,
    which has a smaller volume but higher mass resolution than the \emph{Tiamat} simulation.}
    \end{minipage}
\end{figure*}
In this appendix we present the luminosity--stellar mass relation, and luminosity--halo mass relations based on the
high-resolution \emph{Tiny Tiamat} simulation in order to investigate whether the flattening at low luminosities seen in
Figure \ref{fig:l-mg} and Figure \ref{fig:l-mvir} was due to resolution effects.
Figure \ref{fig:l-mg_tiny} shows the $\log M_*$--$M_\mathrm{UV}$ relation for the model based on \emph{Tiny Tiamat}.
\begin{table}
 \caption{The best fitting slopes and the intercepts of the $\log M_*$--$M_\mathrm{UV}$ relation for galaxies with
          $M_\mathrm{UV}{\leqslant}{-14}$ based on the \emph{Tiny Tiamat} N-body simulation.}\label{tab:l-mg_tiny}
 \centering
 \begin{tabular}{lcc}
  \hline
  \parbox{5mm}{$z$}& \parbox{28mm}{$\mathrm{d}\log M_*/\mathrm{d}M_\mathrm{UV}$}
                   & \parbox{28mm}{$\log M_{* (M_\mathrm{UV}{=}{-19.5})}$}\\
  \hline
       $6$ & $-0.447\pm 0.016$& $8.94\pm 0.05$\\
       $7$ & $-0.424\pm 0.015$& $8.71\pm 0.05$\\
       $8$ & $-0.435\pm 0.029$& $8.63\pm 0.11$\\
       $9$ & $-0.405\pm 0.017$& $8.42\pm 0.06$\\
       $10$ & $-0.437\pm 0.009$& $8.49\pm 0.04$\\
  \hline
 \end{tabular}
\end{table}
We linearly fit the relation for the bright galaxies with $M_{1500}{\lesssim}{-14}$. The best-fit slopes and intercepts are shown
in Table \ref{tab:l-mg_tiny}. The best-fit slopes are larger than the slopes from the model based on \emph{Tiamat}.
The relation becomes flat at $M_\mathrm{1500}{>}{-13}$ which is $1$ magnitude fainter than the model based on \emph{Tiamat}.

\begin{figure*}
    \begin{minipage}{0.99\textwidth}
    \includegraphics[width=\columnwidth]{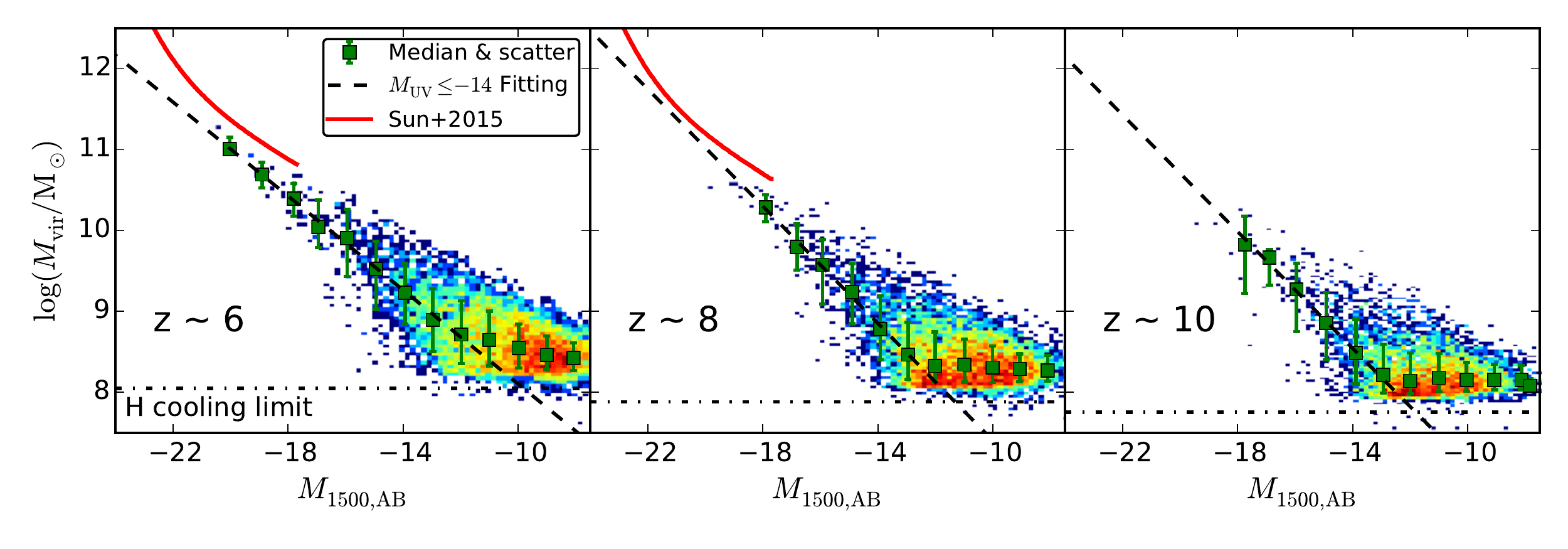}
    \caption{\label{fig:l-mvir_tiny}
    Same as Figure \ref{fig:l-mvir} but for model galaxies at $z{\sim}6$, $8$ and $10$ based on the \emph{Tiny Tiamat} N-body simulation,
    which has a smaller volume but higher mass resolution than the \emph{Tiamat} simulation.}
    \end{minipage}
\end{figure*}
Figure \ref{fig:l-mvir_tiny} shows the halo mass-luminosity relation for all model galaxies based on \emph{Tiny Tiamat}
at $z{\sim}5$--$10$. We fit the $\log(M_\mathrm{vir})$ and the UV magnitudes for galaxies brighter than ${-14}$ mag at each
redshift using a linear relation, as shown by the dashed lines in Figure \ref{fig:l-mvir_tiny}. The values of best-fit
slopes and intercepts are shown in Table \ref{tab:l-mvir_tiny}. The figure clearly shows that the model prevents star
formation in haloes below the hydrogen-cooling limit.
\begin{table}
 \caption{The best fitting slopes and the intercepts of the $\log M_\mathrm{vir}$--$M_\mathrm{UV}$ relation for galaxies with
          $M_\mathrm{UV}{\leqslant}{-14}$ based on the \emph{Tiny Tiamat} N-body simulation.}\label{tab:l-mvir_tiny}
 \centering
 \begin{tabular}{lcc}
  \hline
  \parbox{5mm}{$z$}& \parbox{28mm}{$\mathrm{d}\log M_\mathrm{vir}/\mathrm{d}M_\mathrm{UV}$}
                   & \parbox{28mm}{$\log M_{\mathrm{vir}, (M_\mathrm{UV}{=}{-19.5})}$}\\
  \hline
        $6$ & $-0.290\pm 0.009$& $10.86\pm 0.03$\\
        $7$ & $-0.325\pm 0.018$& $10.84\pm 0.07$\\
        $8$ & $-0.365\pm 0.019$& $10.86\pm 0.07$\\
        $9$ & $-0.319\pm 0.024$& $10.46\pm 0.09$\\
        $10$ & $-0.379\pm 0.006$& $10.61\pm 0.02$\\
  \hline
 \end{tabular}
\end{table}

In summary, the conclusions regarding relations between luminosity and stellar or halo mass are not an artifact of simulation
resolution. We do find that there are small differences in the predicted luminosities within the smallest haloes, owing to the
additional star formation in the merger triggered starburst phase.
\newpage

\bsp

\label{lastpage}

\end{document}